\documentclass[journal]{IEEEtran}

\usepackage{algorithmicx}
\usepackage{algpseudocode}
\usepackage{pseudocode}
\usepackage{algorithm}
\usepackage{epsfig} 
\usepackage{graphicx} 
\usepackage{amsmath} 
\usepackage{amssymb}  
\usepackage{array}
\usepackage{enumerate}
\usepackage{subfig}
\usepackage{float}
\usepackage{cite} 
\usepackage{balance}
\usepackage{amsthm}
\usepackage{stfloats}
\usepackage{epstopdf}
\usepackage{multirow}
\usepackage{tabularx}
\usepackage{ragged2e}
\usepackage{booktabs}
\usepackage{hhline}
\usepackage{lettrine}
\usepackage[dvipsnames]{xcolor}

\usepackage{rotating} 

\usepackage{longtable}

\renewcommand{\arraystretch}{1.0}

\newcolumntype{Y}{>{\centering\arraybackslash}X}

\ifCLASSINFOpdf
  
\else
  
\fi

\begin{document}
\title{Mobility based network lifetime in wireless sensor networks: A review}
\author{Linh Nguyen,~\IEEEmembership{Member,~IEEE} and Hoc T. Nguyen
\thanks{Linh Nguyen is with Faculty of Engineering and Information Technology, University of Technology Sydney, NSW 2007, Australia (e-mail: vanlinh.nguyen@uts.edu.au).

Hoc T. Nguyen is with Department of Networked Systems and Services, Budapest University of Technology and Economics, Budapest 1111, Hungary (e-mail: nguyenthaihoc@vnua.edu.vn or nguyenth@hit.bme.hu).}}


\maketitle

\begin{abstract}
Increasingly emerging technologies in micro-electromechanical systems and wireless communications allows a mobile wireless sensor networks (MWSN) to be a more and more powerful mean in many applications such as habitat and environmental monitoring, traffic observing, battlefield surveillance, smart homes and smart cities. Nevertheless, due to sensor battery constraints, energy-efficiently operating a MWSN is paramount importance in those applications; and a plethora of approaches have been proposed to elongate the network longevity at most possible. Therefore, this paper provides a comprehensive review on the developed methods that exploit mobility of sensor nodes and/or sink(s) to effectively maximize the lifetime of a MWSN. The survey systematically classifies the algorithms into categories where the MWSN is equipped with mobile sensor nodes, one mobile sink or multiple mobile sinks. How to drive the mobile sink(s) for energy efficiency in the network is also fully reviewed and reported.
\end{abstract}



\section{Introduction}
\lettrine[lines=2]{R}{ecently}, given technological advancements in micro-electromechanical
systems and wireless communications, particularly in the Internet of Things (IoT) era, a mobile wireless sensor network (MWSN) \cite{Nguyen2016a,ekici2006mobility} plays a significant impact on in situ observations in variety of environmental and event monitoring applications such as exploring spatial phenomena \cite{Nguyen2017c, Nguyen2019a, Nguyen2014a}, monitoring natural habitats \cite{karl2007protocols}, tracking a target \cite{elhoseny2018optimizing, nguyen2016position}, observing traffic \cite{jain2006exploiting} or battlefield \cite{banerjee2010increasing} and detecting forest fire \cite{alkhatib2014review}. In terms of architecture, mobility in a MWSN can be presented by mobile sensor nodes and/or mobile sink(s). While the mobile sensor nodes are exploited for sensing and/or relaying tasks, the mobile sink(s) is/are employed for gathering the sensed data from the sensor nodes. In some applications, a mobile sink (MS) can also be known as a mobile data collector, which forwards collected data to a base station. In a small-scale network, if a base station traverses around the sensing field for data collection, it can be called as a MS. In contrast to a stationary wireless sensor network (WSN) \cite{Akyildiz2002, levendovszky2014apply}, where both the sensors and base station are firmly located at predefined positions for whole their lives, a MWSN is able to constantly adapt to the changes in the environment and robustly respond to failures of the sensor nodes.

	\begin{figure}[tb]
	\centering
	\includegraphics[width=0.5\textwidth]{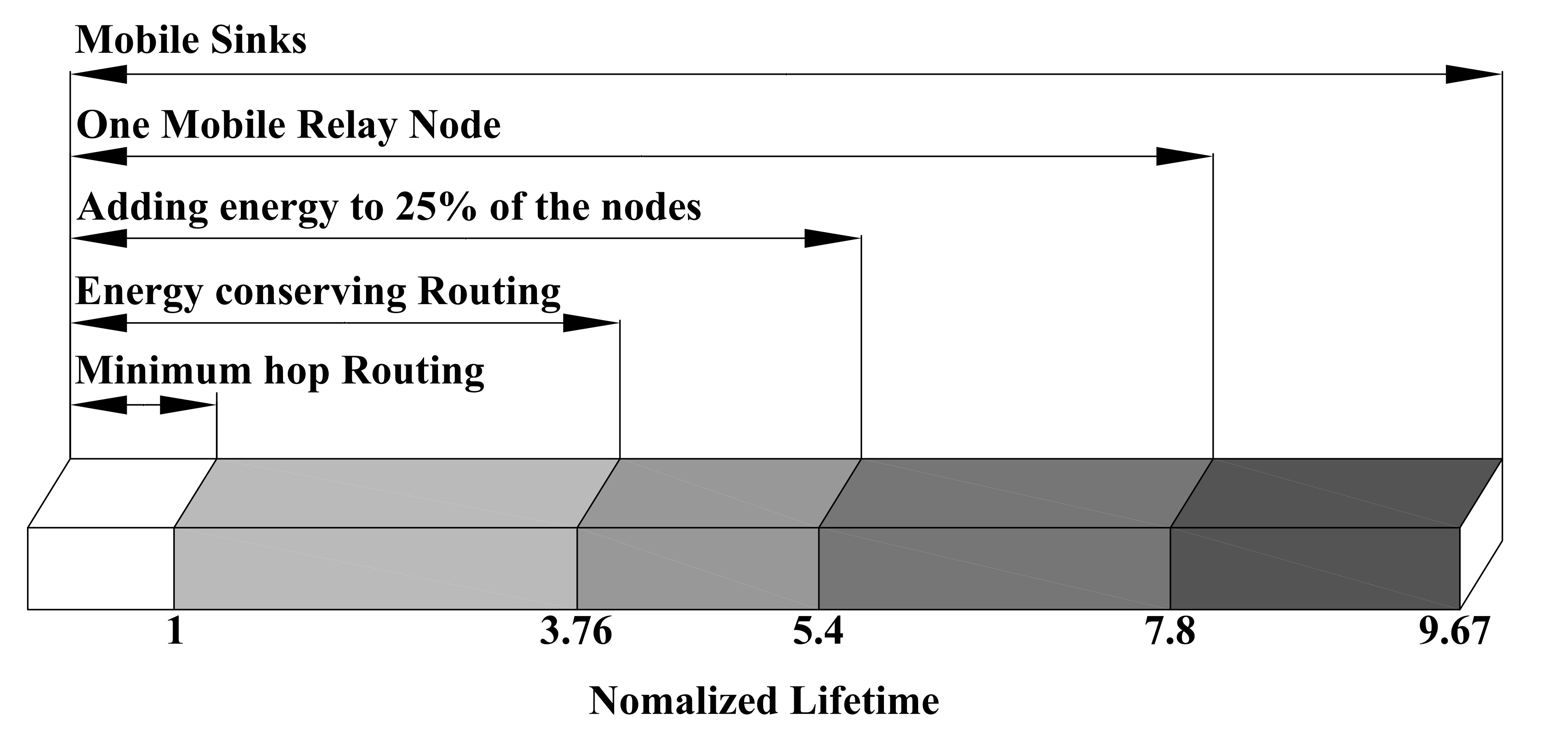}
	\caption{Comparing network lifetime for different approaches in \cite{wang2008extending} }\label{FigF1}
	\end{figure}

In the context of technical performances, mobility enables the sink(s)/nodes to move closer to the transmitters/receivers to reduce transmission distances, which leads a MWSN to better throughput and data fidelity as compared to those in a static WSN \cite{munir2007mobile, kinalis2014biased, kong2014energy, natalizio2013controlled, oudidi2009using}. On the other hand, one of the biggest issues in a stationary WSN is the hot-spot or bottleneck problem, where the closer to the sink the sensor nodes are, the shorter their lifetime is since they have to consume their own energy on transmitting the data of the far-off sensors to the sink. Thus, keeping a sensor node alive over a long time is paramount in the WSN applications as its battery energy is limited and replacing or replenishing that battery is usually impractical. It has been proved that lifetime of a WSN can be effectively elongated if power consumption on each sensor node is minimized and energy load among the network is dexterously balanced \cite{anisi2012overview, yang2010improving}. In literature, many methods have been proposed to extend longevity of a WSN \cite{Nguyen2012b, Nayak2017, Mitici2016, Kacimi2013}; nevertheless, employing mobile platforms is considered as an orthogonal approach to not only address the hot-spot problem but also maximize the lifetime of a WSN \cite{khan2013static, kumar2013sensors, han2016survey, liu2011moving,toumi2016dynamic,verma2014survey,wang2014energy,zhu2015tree}. Interestingly, in many MWSN applications, the mobile elements are naturally available in the sensing field. For instance, animals in habitat monitoring or soldiers in battlefield observing applications can carry the sensors and play as the mobile nodes; moreover, vehicles can be exploited as the mobile sinks (MSs) in a WSN to observe traffic conditions.

More particularly, it is technically noted that wireless communication consumes most of the battery power of each sensor node \cite{ekici2006mobility}. Therefore, if a MS can travel toward sensor nodes, power dissipation on each sensor is substantially reduced. Furthermore, in the scenarios of either node or sink mobility, the sensors can be interchangeably located next to the sink(s), which results in balanced energy consumption over the network \cite{oudidi2009using}. By exploiting mobility in a WSN, the network can be apparently elongated at most as compared with other proposed methods \cite{wang2008extending} as can be seen in Fig. \ref{FigF1}. For instance, by adding one mobile relay node or multiple MSs, the network longevity can be prolonged up to 7.8 or 9.67 times as compared with the minimum hop routing technique, respectively.

Although there is a rich library of approaches proposed to exploit mobility in maximizing the lifetime of a MWSN, to the best of our knowledge it lacks a survey of those techniques. As a result, in this paper we will systematically provide a comprehensive review in that regard. Depending on the type of mobile elements employed in a MWSN, the review will go through the mobility patterns including sensor node mobility, sink mobility and event mobility. The contributions of this paper can be summarized as follows.

\begin{itemize}
\item We first introduce some common concepts broadly utilized to define the lifetime of a MWSN, which allows all readers to appropriately follow the mobility based approaches for energy efficiency discussed in the following sections.

\item We then survey all the techniques using the sensor node mobility for extending the MWSN elongation, where strategies to drive the sensors are based on criteria including coverage, energy-aware, cooperative computing, localization and clustering.

\item In the third contribution, we focus on reviewing the proposed algorithms for prolonging the WSN longevity in which there is only one MS to be utilized. Those algorithms are classified, subject to whether the sink traverses on trajectories in a random, fixed/predictable or controlled manner.

\item Akin to the third contribution, we explore and then comprehensively summarize the approaches employed in a MWSN with multiple MSs.

\item Last but not least, what methods developed for energy efficiency in a MWSN for tracking an event, where objects or targets are mobile and arbitrarily appear, are systematically surveyed.
\end{itemize}

The remaining of the paper is arranged as follows. Section \ref{sec_2} presents some widely used definitions of the WSN lifetime before Section \ref{sec_3} summarizes the sensor node mobility based energy-efficient algorithms in a MWSN. Section \ref{sec_4} and Section \ref{sec_5} review how to randomly, predictably and controllably drive a MS and multiple MSs in an energy-efficient oriented MWSN, respectively. Event mobility based approaches are surveyed in Section \ref{sec_6} before conclusions are drawn in Section \ref{sec_7}.

\section{Wireless Sensor Network Lifetime Definitions}
\label{sec_2}
Lifetime of a WSN can be understood as the total amount of time from the network's initial deployment until the network's incapability of responding a sensing requirement or archiving a particular objective. It is a critical criterion in designing, operating and maintaining the network. Due to limitation of an individual sensor's battery, which is usually impractical to be recharged or replaced, many works have been conducted to ameliorate the network lifetime. Nonetheless, depending on different specific sensing tasks or objective functions, definitions of the WSN lifetime can specifically vary from an application to another application. For a quite comprehensive summary of those definitions, interested readers may be referred to \cite{Dietrich2009}; nevertheless, in this section, we introduce some common network lifetime concepts so that all readers can conveniently follow mobility based techniques employed to enhance the lifetime of a WSN, which will be discussed in the following sections.

Mathematically, the lifetime of a WSN can be generally computed as follows \cite{chen2005lifetime,blough2002investigating},
\begin{equation}\label{eqgenerallifetime}
\mathbb{E}\left[ L \right] = \frac{{{\varepsilon _0} - \mathbb{E}\left[ {{E_\omega }} \right]}}{{{P_c} + \lambda \mathbb{E}\left[ {{E_r}} \right]}},
\end{equation}
where $\mathbb{E}\left[ L \right]$ is the expected average lifetime of the network. In terms of the network-wise, ${{\varepsilon _0}}$ denotes the total non-rechargeable initial battery power, and $P_c$ defines the total constant continuous energy expensed. While $\lambda$ is the number of data collection times every time unit, both $\mathbb{E}\left[ {{E_\omega }} \right]$ and $\mathbb{E}\left[ {{E_r}} \right]$ represent the total unused energy and the total consumed energy over the network, respectively. The equation (\ref{eqgenerallifetime}) can be employed to work out how long a WSN can serve for a sensing application regardless its underlying network model factors such as network architecture and protocol, data collection initiation and channel fading characteristics.

In a very simple definition, the lifetime of a WSN can be heuristically deemed as the total time the network can operate in a sensing task until the first sensor node dies \cite{agyei2013lifetime,levendovszky2010fading}. Based on this concept, the work in \cite{kang2003maximizing} proposed a max-min type optimization approach to prolong the network longevity by maximizing the working time of the first sensor node. However, in some applications, the remaining active sensor nodes after the first one died can still provide appropriate functionalities \cite{najimi2014lifetime}, which makes the first dead sensor node based network lifetime definition too pessimistic. As a result, the authors in \cite{najimi2014lifetime,key:article} called a WSN to be dead if a certain percentage of the total sensors dies; that is, balanced load among the sensor nodes may significantly lead to elongating the network longevity. In the most optimistic point of view, Khan \textit{et al.} in \cite{khan2015dyn} considered that the network is still able to provide useful services until the last sensor dies though this definition is hardly applied for realistic applications, where coverage is a crucial parameter. 

\begin{table*}[tb]
\renewcommand{\arraystretch}{1.3}
	\caption{ADVANTAGES AND DISADVANTAGES OF NETWORK LIFETIME DEFINITIONS}
	\label{table1}
	\centering
		\begin{tabular}{|>{\arraybackslash}m{1.4cm}|>{\arraybackslash}m{3.6cm}|>{\arraybackslash}m{5.6cm}|>{\arraybackslash}m{5.6cm}|} \hline
\multicolumn{2}{|l|}{\textbf{Categories of definition}}& \textbf{Advantages} & \textbf{Disadvantages}\\ \hline
\multirow{11}{1.4cm}{\textbf {Number of reliable nodes}}&
			Network lifetime is defined as amount of time from the network’s initial deployment to the first node fails \cite{agyei2013lifetime,levendovszky2010fading}. & It is straightforward to compute network lifetime. And the definition can be employed in a max-min type optimization problem where longevity of the first node is maximized \cite{kang2003maximizing}. & The definition  seems too pessimistic since it is very likely that when the first node fails, the rest of the nodes still can provide appropriate functionality \cite{najimi2014lifetime}. It can be only reasonably used if all sensor nodes have a similar energy consumption rate. The coverage and connectivity issues in WSNs are not considered in the definition. \\ \cline{2-4}
			&  The lifetime of a sensor network is defined as total time in which a certain percentage of total nodes is drained of energy 
			\cite{najimi2014lifetime,key:article}. & The definition is flexible in applications where a percentage threshold can be prescribed. The network lifetime can be significantly enhanced if load among nodes is balanced. & Longevity of a network is not accurately calculated when there is uniform distribution of dead nodes in the network. Moreover, other crucial performance indices such as coverage and connectivity are not included in the definition. \\ \cline{2-4}
			& The network lifetime is defined as total working time of the network until its all nodes run out of their energy \cite{khan2015dyn}. & The definition is straightforward. & This type of definition is rarely employed and can be inappropriate for practical applications since a network may stop providing useful services far from failure of the last node. Coverage and connectivity in the network are not incorporated into the formula.\\ \hline
			\multirow{8}{1.4cm}{\textbf {Coverage}}&
			 Area coverage (each location of the interest area is monitored by at least one sensor node) \cite{khan2015dyn}. & The definition can lead to effective strategies for monitoring applications in WSNs. & Monitoring every single point in a sensing field is too stringent in some applications where it is not necessary to observe the whole area.\\ \cline{2-4}
			&Target coverage (a network monitors a fixed number of targets) \cite{idrees2015distributed,liu2011moving}. & A predefined coverage threshold can be reduced to zero when all sensors in a network run out of energy \cite{khan2015dyn}. & Ability to sense events of sensor nodes is not indicated.\\ \cline{2-4}
			&Barrier coverage (probability of some moving targets, which are observed by sensor nodes when they pass through a sensing field \cite{cardei2005energy}. & A k-discrete barrier coverage model can be employed to deploy sensors to form $k-$lines barriers \cite{du2013maximizing}. & Communication among sensor nodes in a network is not considered in the definition.\\ \hline
			
			\multicolumn{2}{|c|}{\textbf {Communication connectivity}} & Communication connectivity is one of the most critical tasks of a sensor node. Thus, a network lifetime definition considering a maximum number of communication rounds is frequently utilized \cite{hajiaghayi2010maximizing,legakis2015lifetime,blough2002investigating,degirmenci2014maximizing}. & Other performance indices including coverage or quality of service are not embedded in the definition.\\
			\hline      
			\multicolumn{2}{|c|}{\textbf {General formula \cite{chen2005lifetime}}} & Longevity of a network can be significantly improved while wasted energy (i.e., total unused energy in the network when it is dead) is reduced. & The definition is not easy to be implemented due to its complexity and undetermined characteristics.\\
			\hline                                        
			\multicolumn{2}{|c|}{\textbf {Application requirements \cite{kumar2005lifetime,li2009network}}} & It is subject to application designs. & In some applications, where there are concurrent requirements such as coverage degree and notification latency, it is complicated to incorporate those requirements into the definition.\\  \hline
		\end{tabular}
\end{table*}

In the context of coverage, elongation of a WSN can be considered as the total amount of time during which the network provides full coverage in an expected region. If there appears any hole in the covered region, which is not monitored by any sensor node, the WSN is dead \cite{liao2011ant, idrees2015distributed}. Categorically, there are three types of coverage required to be covered by a WSN in realistic applications: (i) area coverage, e.g. every single point in an interested area is observed by at least one sensor node \cite{khan2015dyn}, (ii) target coverage, e.g. a fixed number of targets are constantly tracked \cite{idrees2015distributed,liu2011moving} and (iii) barrier coverage, e.g. probability of moving targets randomly passing through a sensing field, captured by sensor nodes \cite{cardei2005energy, du2013maximizing}. In conjunction with data transmission, the authors in \cite{kim2013lifetime} further refined the definition of the network lifetime by adding more requirement. That is, the network is considered to be dead if sensed data cannot be transmitted to the base station though the sensors are still able to provide sensing. In that case, even though the sensor nodes are still alive but they are isolated or disconnected from the others. As a consequence, some works \cite{hajiaghayi2010maximizing, legakis2015lifetime, degirmenci2014maximizing} utilized communication criteria to define the lifetime of a WSN. For instance, Hajiaghayi \textit{et al.} in \cite{hajiaghayi2010maximizing} considered that the network is still useful if it can guarantee a minimum data transfer rate. Under consideration of other quality of service parameters such as packet loss or time delay, the work \cite{legakis2015lifetime} suggested that a WSN can provide full functionalities until its probability of connectivity goes below a predefined threshold. Nevertheless, as pointed out by \cite{blough2002investigating}, the quality of communication is not always a sufficient criterion to determine the longevity of a WSN. Thus, the authors in \cite{degirmenci2014maximizing} proposed another definition based on query, which is dependent on both connectivity and quality of service constraints such as transmission range, time-to-live counter and active/sleep schedules. In a stricter term, since a network is simultaneously affected by many factors such energy consumption, coverage, connectivity and so on, some researchers employed both coverage and connectivity in a criterion to better and more accurately identify the network elongation \cite{guimaraes2016increasing,liao2015energy,mo2005mostly}.
 
In some special scenarios, the network lifetime is truly dependent on specific requirements of applications including degree of coverage of latency of notification \cite{kumar2005lifetime,li2009network}. For instance, in the work \cite{li2009network} Li \textit{et al.} employed a number of rounds of estimation the network can archive before it becomes not completely functional as a definition of the network longevity.

Table \ref{table1} categorizes the widely used network lifetime definitions, which also briefly provides advantages and disadvantages of each approach.


\section{Sensor Node Mobility for Efficient Energy}
\label{sec_3}
The significant benefit of the MWSN as compared with a static WSN is that limited energy of the batteries can be efficiently and more equally consumed among the sensor nodes. Let us consider a static WSN, where all the sensor nodes are stationary after being deployed. It is noted that the sensor nodes closer to the sink have to transmit not only their own data but also the information packages from the other nodes further away from the sink to the sink. When the sensor nodes forward the data of the others, they play as a relay node \cite{wang2008extending} in those multi-hop communications. Hence, it can be apparently seen that energy of the nodes nearby the base station is more quickly depleted. This problem is known as the bottleneck or hot-spot issue in the network, which ultimately leads to shortage of the network lifetime. In contrast, due to the mobile ability, the sensor nodes in the MWSN can consume their energy more flexibly and efficiently \cite{munir2007mobile}. Many schemes have been proposed to effectively reduce energy consumption in the MWSN as discussed in the following.

	\begin{figure}[tb]
	\centering
	\includegraphics[width=0.5\textwidth]{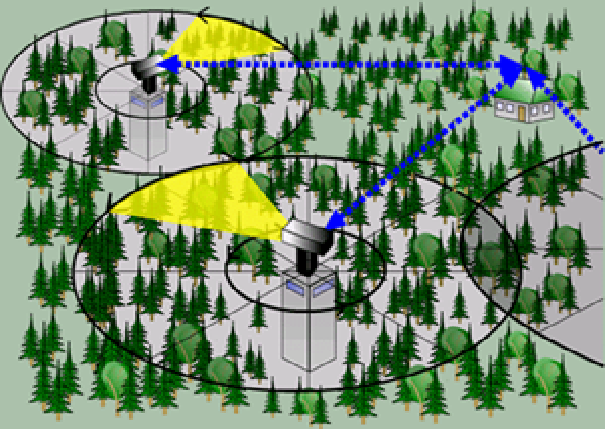}
	\caption{Improving connectivity and coverage in the forest fire detection  \cite{alkhatib2014review} }\label{FigF2}
\end{figure}
	
\subsection{Coverage based Strategies}
By taking advantage of movements, the WSN with mobile sensor platforms has been exploited in many monitoring applications, e.g. detecting forest fire \cite{alkhatib2014review} as shown in Fig. \ref{FigF2}, where it is required to effectively cover an area of interest while well maintaining connectivity throughout the network. In other words, mobility allows the sensor nodes to be able to change their locations after being initially deployed. The coverage and connectivity coupling problems have been thoroughly studied in the literature \cite{mohamed2017coverage,romer2004design}. For instance, some works \cite{liu2011moving,yang2007movement} focused on mobilizing the initial deployment of the network in order to change the topology of the network for better connectivity and coverage. In \cite{yang2007movement} Yang \textit{et al.} proposed a sensor deployment strategy for the MWSN, where in order to reduce the energy holes near the sink and prolong the network lifetime, the desired non-uniform sensor density in the monitored area is formed. By assuming that each sensor can move only once, a centralized maximum-flow minimum-cost algorithm is then formulated to relocate the mobile sensors so that the sensor density is guaranteed given a minimum energy consumption. By dividing the interested area into $n$ coronas in which all the sensor nodes have the same energy consumption rate, the proposed approach has demonstrated that the lifetime of a network with a non-uniform sensor distribution can be improved at least $n$ time higher than that of a network with a uniform sensor distribution. Nonetheless, the centralized technique is limited in a large distributed wireless sensor network \cite{liao2011ant} and inefficient when energy of all the sensor nodes in the network is at different levels. In a similar fashion, the authors in \cite{liu2011moving} proposed another algorithm, called Moving Algorithm for Achieving the Non-uniform Deployment (MAND) for a non-uniform distributed network. By employing Voronoi diagram technique \cite{ab2009wireless}, the proposed method rearranges the mobile sensors at an appropriate distance with their neighbor peers so that their desired density is guaranteed and the coverage holes \cite{luo2009double}  are eliminated, which, on the other hand, significantly reduces the energy consumption in the network. An extended version of the algorithm called EMAND was also considered, where the sensor nodes tend to move to areas with the sparser density. The proposed technique is considerably beneficial to applications monitoring environments with high event generating rate. Yet, it cannot guarantee complete coverage of the monitored areas \cite{chen2005lifetime}.

Akin to the previous work relying on the Voronoi diagram method, Abo-Zahhad \textit{et al.} in \cite{abo2016centralized} incorporated the muli-objective immune technique \cite{abo2014coverage} into Voronoi cells in order to enhance both coverage and longevity of the MWSN. The proposed algorithm called Centralized Immune-Voronoi deployment Algorithm (CIVA) allows the MWSN to simultaneously manage mobility, sensing range and active/sleep mode of each mobile sensor. In other words, it first computes an optimal number of the sensors needed to complete coverage and then relocate those mobile platforms in the monitored area so that energy consumption during movement and sensing is minimized. Furthermore, the work also studied the coverage issue in the MWSN under obstacles given both binary and probabilistic models. In the work \cite{liao2015minimizing} the authors paid more attention on minimizing energy consumed during the mobile sensor movements to improve the lifetime of the network using for tracking targets. Again, the Voronoi diagram was exploited to create clusters of the sensors depending on their vicinity to the targets. The concept of the Voronoi diagram of the targets was first adopted in the application, which leads to reduction of both computational complexity and repeatability.

It is noticed that both scheduling for coverage and maximizing lifetime in the WSN are NP-hard problems \cite{articlePark,zhou2004connected}. Therefore, the optimal solution for the maximum coverage and lifetime coupling problem cannot be found in a polynomial time. There are some heuristic algorithms proposed to find its near-optimal counterpart. For instance, Ebrahimnezhad \textit{et al.} in \cite{articleEbrahimnezhad2011} proposed the Improved Harmony Search (IHS) algorithm to search for a near-optimal solution for the coverage and lifetime of the $k$-coverage WSN. The premise behind the proposed method is to deploy the sensor nodes in a geometrically symmetric and circle-like shape and position the sink in its centre, which enables all the other sensors to have the shortest communication path and balance load among them. More importantly, given mobility, some sensor nodes consumed high energy are relocated at the closest positions of the others consumed low energy and vice versa after a time. This proposition results in enhancing the longevity of the MWSN while both coverage and connectivity are guaranteed.
Two other heuristic algorithms based on the genetic and ant colony optimization were also proposed in \cite{elhoseny2018optimizing,liao2011ant,katsuma2008maximizing,farzana2017ant} to address the sensor deployment problem in a network of wireless sensors, aiming to maximize the coverage in the WSN for continuously monitoring a specific region in the longest possible period of time.

\begin{table*}
\caption{TECHNIQUES FOR IMPROVING NETWORK LIFETIME BASED ON SENSOR NODE MOBILITY}
\begin{center}
\begin{tabular}{m{0.25cm} m{0.25cm} m{0.25cm} m{5cm} m{0.25cm} m{0.25cm} m{0.25cm} m{5cm}}
	\hline
\multicolumn{1}{c} {\bf A1}&\multicolumn{1}{c} {\bf A2}&\multicolumn{1}{c} {\bf A3}&\multicolumn{1}{c} {\bf Objectives and Methods}&\multicolumn{1}{c} {\bf A4}&\multicolumn{1}{c} {\bf A5}&\multicolumn{1}{c} {\bf A6}&\multicolumn{1}{c} {\bf Issues handled}\\
	\hline
\multicolumn{8}{l}{\bf A. Coverage based strategies} \\ 
\cite{liu2011moving} & \multicolumn{1}{c}{Yes} & \multicolumn{1}{c}{Flat} & Moving mobile sensors to appropriate locations for better coverage and prolong network lifetime. & \multicolumn{1}{c}{Yes} & \multicolumn{1}{c}{Yes} & \multicolumn{1}{c}{Yes} & Improving connectivity and coverage and prolonging network lifetime. \\ 
\cite{yang2007movement} & \multicolumn{1}{c}{No} & \multicolumn{1}{c}{Cluster} & Relocating mobile sensors by a centralized scheme in order to achieve density requirement. & \multicolumn{1}{c}{Yes} & \multicolumn{1}{c}{NA} & \multicolumn{1}{c}{Yes} & Improving coverage, prolonging the network lifetime and reducing energy hole. \\ 
\cite{liao2011ant} & \multicolumn{1}{c}{No} & \multicolumn{1}{c}{Cluster} & Modelling the sensor deployment problem as the multiple knapsack problem to extend network lifetime while still maintain sensing coverage. & \multicolumn{1}{c}{NA} & \multicolumn{1}{c}{NA} & \multicolumn{1}{c}{NA} & Improving coverage and prolonging network lifetime. \\ 
\cite{ab2009wireless} & \multicolumn{1}{c}{Yes} & \multicolumn{1}{c}{Flat} & Finding the optimal locations of sensor nodes based on Voronoi diagram & \multicolumn{1}{c}{NA} & \multicolumn{1}{c}{NA} & \multicolumn{1}{c}{NA} & Providing good coverage within a reasonable computational time \\ 
\cite{luo2009double}& \multicolumn{1}{c}{Yes} & \multicolumn{1}{c}{Flat} & Considering interaction between uncontrollable mobility and controllable mobility when deploying sensors to improve the coverage issue. & \multicolumn{1}{c}{Yes} & \multicolumn{1}{c}{Yes} & \multicolumn{1}{c}{NA} & Extending system lifetime and improving network coverage. \\ 
\cite{abo2014coverage}& \multicolumn{1}{c}{Yes} & \multicolumn{1}{c}{Flat} & Minimizing moving dissipated energy so that mobile sensor nodes will be relocated for better coverage. & \multicolumn{1}{c}{Yes} & \multicolumn{1}{c}{NA} & \multicolumn{1}{c}{NA} & Maximizing covered area and minimizing moving energy consumption. \\ 
\cite{liao2015minimizing} & \multicolumn{1}{c}{Yes} & \multicolumn{1}{c}{Flat} & Deploying mobile sensor nodes in a network with minimum movement. & \multicolumn{1}{c}{Yes} & \multicolumn{1}{c}{NA} & \multicolumn{1}{c}{Yes} & Providing target coverage and network connectivity with requirements of moving sensors. \\ 
\multicolumn{8}{l}{\bf B. Energy - aware based strategies} \\ 
\cite{el2014energy} & \multicolumn{1}{c}{Yes} & \multicolumn{1}{c}{Cluster} & Relocating a minimum number of redundant nodes towards center of interest (COI) while still maintaining connectivity between COI and region of interest (ROI). & \multicolumn{1}{c}{Yes} & \multicolumn{1}{c}{Yes} & \multicolumn{1}{c}{Yes} & Reducing energy consumption of sensor nodes consumed for their relocation tasks and rerouteing between COI and ROI when a failed node occurs. \\ 
\cite{Rao2004} & \multicolumn{1}{c}{Yes} & \multicolumn{1}{c}{Flat} & Driving mobile sensor nodes through environment relied on a distributed simulate annealing framework. & \multicolumn{1}{c}{Yes} & \multicolumn{1}{c}{Yes} & \multicolumn{1}{c}{NA} & Minimizing expensed network power in both mobility and communication.  \\ 
\cite{Wang2005} & \multicolumn{1}{c}{Yes} & \multicolumn{1}{c}{Flat} & Evaluating network lifetime in three scenarios: (1) when the network is all static, (2) when the network has one mobile sink and (3) when the network has one mobile relay. & \multicolumn{1}{c}{Yes} & \multicolumn{1}{c}{Yes} & \multicolumn{1}{c}{Yes} & Improving network lifetime and mitigating the bottleneck problem.  \\ 
\multicolumn{8}{l}{\bf C. Cooperative computing based strategies} \\ 
\cite{sheng2018energy} & \multicolumn{1}{c}{No} & \multicolumn{1}{c}{Flat} & Minimizing energy consumption of process in MWSNs while fulfilling requirements of completion time. & \multicolumn{1}{c}{No} & \multicolumn{1}{c}{No} & \multicolumn{1}{c}{NA} & Decreasing total energy consumption while ensuring a given level of completion time. \\ 
\multicolumn{8}{l}{\bf D. Localization based strategies } \\ 
\cite{hu2004localization}& \multicolumn{1}{c}{No} & \multicolumn{1}{c}{Flat} & Locating positions of sensor nodes by using the sequential Monte Carlo localization method. & \multicolumn{1}{c}{Yes} & \multicolumn{1}{c}{Yes} & \multicolumn{1}{c}{Yes} & Improving accuracy and reducing costs of localization. \\ 
\cite{baggio2008monte} & \multicolumn{1}{c}{No} & \multicolumn{1}{c}{Flat} & Using the sequential Monte Carlo localization boxed to locate positions of sensor nodes. & \multicolumn{1}{c}{NA} & \multicolumn{1}{c}{Yes} & \multicolumn{1}{c}{Yes} & Improving accuracy of localization. \\ 
\cite{idris2017low} & \multicolumn{1}{c}{No} & \multicolumn{1}{c}{Flat} & Locating positions of sensor nodes by two Monte Carlo based localization schemes termed MCL (Monte Carlo localization) and MSL (mobile and static sensor network localization). & \multicolumn{1}{c}{Yes} & \multicolumn{1}{c}{Yes} & \multicolumn{1}{c}{No} & Improving accuracy of localization and reducing  communication costs. \\ 
\cite{Somasundara2006} & \multicolumn{1}{c}{No} & \multicolumn{1}{c}{Cluster} & Controlling motions of mobile sensor nodes to adjust speed of sensors moving on closed trajectories. & \multicolumn{1}{c}{Yes} & \multicolumn{1}{c}{Yes} & \multicolumn{1}{c}{Yes} & Improving network lifetime by improving energy performance. \\ 
\multicolumn{8}{l}{\bf E. Clustering based strategies } \\ 
\cite{7876282} & \multicolumn{1}{c}{No} & \multicolumn{1}{c}{Cluster} & Prolonging network lifetime by the LEACH-CCH cluster algorithm. & \multicolumn{1}{c}{Yes} & \multicolumn{1}{c}{No} & \multicolumn{1}{c}{No} & Improving network lifetime. \\ 
\cite{elhoseny2018optimizing} & \multicolumn{1}{c}{No} & \multicolumn{1}{c}{Cluster} & Tracking a target by the genetic algorithm based coverage approach & \multicolumn{1}{c}{No} & \multicolumn{1}{c}{No} & \multicolumn{1}{c}{No} & Minimizing energy consumption in the network. \\ 
\multicolumn{8}{l}{\it A1: References; A2: Sensor location known; A3: Network structure; A4: Moving distance minimized;} \\ 
\multicolumn{8}{l}{\it A5: Hop-routing minimized; A6: Connectivity required; NA: Not Applicable.} \\ 
\hline
\end{tabular}
\end{center}
\label{table2}
\end{table*}

\subsection{Energy-Aware based Strategies}
Another idea to prolong the lifetime of the WSN is to get information of energy left on each sensor node known. For instance, in the works \cite{el2014energy,el2012achain}, it discusses that after being initially deployed to monitor a region of interest (ROI), the sensor nodes may need to be relocated outside the ROI to cover other events, called centre of interest(COI). In order to sufficiently monitor the events with a minimum number of the sensor nodes while maintaining the connectivity between the ROI and the COI, the authors proposed a new algorithm known as Chain Based Relocation Approach (CBRA). To reduce energy consumption in this exercise, the proposed method first estimate total energy one sensor node would consume if it is moved toward the COI, and then effectively select a minimum number of the sensors to cover the events. In case there is failure occurring in one of the sensors sent to the COI, which leads to disconnecting between the ROI and the COI, the algorithm comprises a fault tolerance procedure to recover the network from errors. In a large-scale network, Rao \textit{et al.} in \cite{Rao2004} suggested a distributed simulate annealing framework to effectively drive the mobile sensor nodes through environment, where they can play as a relay or a relaying and sensing device, so that the network power in both mobility and communication is minimized. It is noticed that in the proposed distributed approach, each mobile sensor only requires local energy information to decide where to go in every step. In the same manner, the work \cite{Wang2005} investigated impact of the mobile nodes in a large sensor network when they act as a relay. It then proved that the network lifetime in a network with mobility can be enhanced by a factor of four as compared with that in a static one.

In the context of uncontrollable mobile wireless sensors, to prolong the network's lifetime, a energy-aware scheme is designed to compute how much utility received and how many sensor nodes alive after each mission \cite{articleRowaihy}. In equivalent words, each sensor node is to be considered its energy level before contributing the measured information to the mission. Expected utility contribution to each mission must be greater than a threshold $r$, which depends on the sensor's fraction of remaining energy $f$. $r$ can be computed by
\begin{equation}
r=\tau^f,
\end{equation}
where $\tau$ is a certain sensing threshold defined by user based on the level of the sensory information and quality the user expects. The proposed algorithm allows the WSN to avoid the less useful missions when the energy levels of the sensor nodes are low, which ultimately extends the longevity of the network.

\subsection{Cooperative Computing based Strategies}
\begin{figure}[tb]
	\centering
	\includegraphics[width=0.5\textwidth]{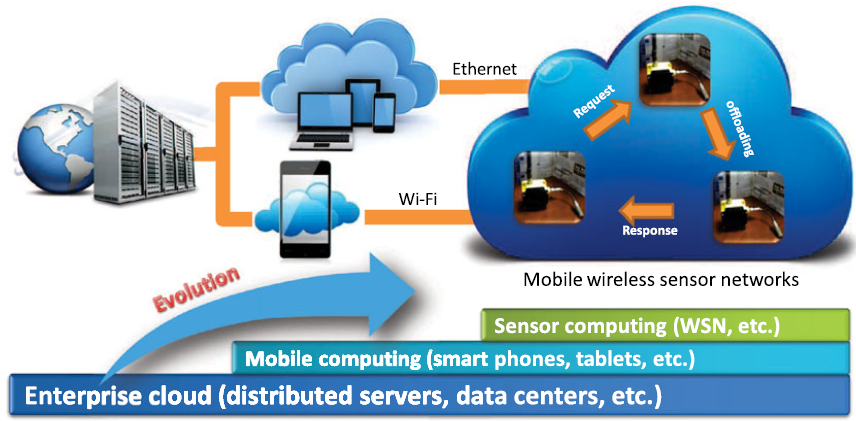}
	\caption{Large distributed sensor-as-a-service infrastructure \cite{sheng2018energy}}\label{WMSN_IoT}
\end{figure}

In era of IoT, the lifetime of the MWSN can be prolonged by employing the cloud computing infrastructure as shown in Fig. \ref{WMSN_IoT}. In other words, the authors in \cite{sheng2018energy} introduced a new cooperative computing approach to boost energy efficiency in a network of mobile sensors, where total energy consumption in processing an application is optimized, given a certain completion deadline. In the proposed method, the sensor nodes are encouraged to share their resources cooperatively so that both computation and communication costs are considered as a whole; and the workload can be optimally partitioned, offloaded and executed among the sensor nodes. Furthermore, the energy efficient strategies are also cooperated for resource allocation in the MWSN, which exploits mobility feature of the sensor nodes to guarantee energy effectiveness given minimum transmission time.

Total energy consumption in the WMSN for the purpose of processing an application given a certain completion deadline $T$ can be minimized by using cooperative computing as follows,
\begin{equation}
\label{first_cri2}
\underset{\begin{array}{c}
l_l+l_r=L\\ t\leq T
\end{array}
}{\operatorname{\textbf{min}}}\; E_c^l(l_l,t)+E_t(l_r,g_{l,r})+E_r(l_r,g_{r,l})+E_c^r(l_r,t),
\end{equation}
where $E_c^l$ and $E_c^r$ are the local and remote computation energy consumption, while $E_t$ and $E_r$ are transmission and reception energy consumption, respectively. $l_l$ and $l_r$ are the partitioned input data sizes for local and remote processing, repetitively. $g_{l,r}=g_{r,l}$ is the channel gain, representing a symmetric channel between the local and remote sensor nodes.

\subsection{Localization based Strategies}
One of aspects consuming energy from the sensor nodes is localizing themselves since in some applications the collected data is meaningless without sensor location information \cite{Nguyen2014b, Nguyen2019b, Nguyen2018b, Nguyen2017a, Nguyen2016b, Nguyen2016c}. Localizing mobile sensors in the WSN can be conducted via the Global Positioning System (GPS) \cite{hofmann2012global}; however, the GPS signals are restricted in indoor, underwater or foliage environments. Range-based or range-free technologies \cite{poellabauer2014range} can also be employed to identify the sensor locations though additional hardware is required, which may ultimately be costly and apparently consume more energy. Other schemes for the sensor network localization do not rely on additional hardware but are based on the Monte Carlo theory were proposed, including Monte Carlo localization \cite{hu2004localization}, Monte Carlo localization boxed \cite{baggio2008monte} and mobile and static sensor network localization \cite{rudafshani2007localization}; nonetheless, they increase communication energy consumption in the network \cite{idris2017low}. In \cite{idris2017low}, a low communication cost (LCC) algorithm was proposed to energy-efficiently localizing the sensor nodes in the MWSN. The proposed approach can estimate the sensor location based on not only anchor nodes but also normal nodes, which are selected from the first and second hop neighbors of the estimated node. In other words, the LCC method aims to minimize dependence of the sensor location estimation procedure on anchor nodes, which practically reduce communication energy consumption in the MWSN. Furthermore, an adaptive approach was proposed in \cite{Somasundara2006} to control motions of mobile sensor nodes, which primarily aims to adjust speed of the sensors moving on the closed trajectories so that they can maximally deliver data within a specified latency constraint. It is noticed that cost of increased data latency may lead to reduction of the WSN lifetime in many applications.

\subsection{Clustering based Strategies}
In this last subsection of employing mobility nature in the MWSN to extend the sensor battery lifetime, we summarize techniques based on clustering the mobile sensors. For instance, in the work \cite{el2014energy}, some of the mobile sensor nodes are sent to the COI to cover the events. In the meantime, the remaining nodes in the ROI are considered whether they are redundant or not by utilizing the clustering technique. Specifically, the sensors in the ROI are clustered into square grid cells; and in each cell, a particular sensor node is elected, relied on its energy level, to be a cell head that is responsible to monitor such a grid square. All other sensor nodes in that cell are switched to the passive mode, i.e. no sensing tasks required. It is obvious that the proposed strategy can considerably decline the energy consumption in the MWSN. In similar manner, Corn \textit{et al.} in \cite{7876282} proposed the Low-Energy Adaptive Clustering Hierarchy Centered Cluster-head (LEACH - CCH) algorithm in order to extend the lifetime of the MWSN. When the mobile sensor nodes are moving around, the proposed method estimates the sensor locations and reconstruct sensor clusters accordingly. As soon as a cluster of the sensor nodes is established, a particular node closest on average to the centre of the cluster is elected as a cluster head, which improves transmission distances for all the non-cluster-head nodes and reduces energy consumed in the network.

Moreover, other works exploited heuristic methods based on the ant colony optimization (ACO) \cite{liao2011ant} and genetic algorithm (GA) \cite{elhoseny2018optimizing} to prolong the longevity of the MWSN. More specifically, in \cite{liao2011ant}, in the ACO based sensor deployment strategy, the authors proposed to cluster the sensor nodes in the service region into the circle points. In each circle point, one sensor node is elected as a sensor head to look after movements of all the others. In equivalent words, the sensor head gathers information about consumed and remaining energy on all the sensor nodes in its circle point and then periodically exchange those data with the counterparts in the neighbor. Given the collected information, the proposed algorithm moves the sensor nodes to the appropriate circle points in order to balance the energy consumption in the whole network, which leads to its longer lifetime. Likewise, in \cite{elhoseny2018optimizing}, Elhoseny \textit{et al.} proposed the GA based coverage approach for tracking target. After moving the mobile sensors around the target field, based on coverage range of each sensor, estimated energy consumption, distances from the sensor nodes to the sink and target positions, the algorithm optimizes a number of the cover heads that are employed to transmit data to the sink. By utilizing a minimum number of the sensor nodes but still guaranteeing coverage of all the targets, the energy consumed by the MWSN is minimized.

\begin{figure}[tb]
	\centering
	\includegraphics[width=0.45\textwidth]{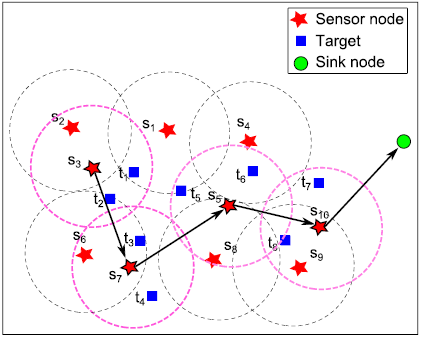}
	\caption{Mobile wireless sensor network for target tracking \cite{elhoseny2018optimizing}}\label{Figcover_target}
\end{figure}

It can be seen in Table \ref{table2}, which provides summarises of the existing methods, that the sensor node mobility plays a significant role in elongating lifetime of a MWSN. Subject to a specific application, the mobile nodes can be deployed in different fashions. It ranges from requiring full coverage in forest fire and earthquake detections, e.g. Fig. \ref{Figcover_target} and target tracking to observing local areas, e.g. battlefield monitoring, where density of the sensor nodes is arranged based on requirements of data transmission. One of limitations of a sensor node is memory and power burden. Hence, another paradigm was proposed in a MWSN is drive a sink \cite{ghafoor2014efficient,han2017disaster,kaswan2018multi,kumar2018aco,mitra2018proactive,xie2017energy,yarinezhad2018reducing} or sinks \cite{gao2011efficient,keskin2017column,kinalis2007scalable,lee2018active,yang2017practical} around the network to collect data from the sensor nodes, which will be discussed in the following sections.

\section{One Sink Mobility based Efficient Energy}
\label{sec_4}
In contrast with the schemes presented in Section \ref{sec_3}, where the sensor nodes are presumed to be mobile, this section investigates taxonomy of the proposed methods that exploit movements of a MS to reduce energy consumption in a MWSN. Subject to the network configuration, allowed data latency, environmental conditions (e.g. terrain, roads and sensing field) and application requirements, the MS can be driven by the random, predictable or fixed-path and controlled strategies \cite{chatzigiannakis2008efficient,tunca2014distributed, nguyenhoc2017efficient}. Brief summaries of the frequently utilized approaches are provided in Table \ref{table3}, and details of those methods are discussed in the following subsections.

\begin{table*}
	\caption{TECHNIQUES FOR IMPROVING NETWORK LIFETIME BASED ON ONE SINK MOBILITY}
	\begin{center}
		\begin{tabular}	{m{0.6cm}m{0.3cm}m{3.5cm} m{0.25cm} m{0.25cm} m{3.5cm} m{3.5cm}}
			\hline
			\multicolumn{1}{c}{\bf A1} & \multicolumn{1}{c}{\bf A2} & \multicolumn{1}{c}{\bf A3} & \multicolumn{1}{c}{\bf A4} & \multicolumn{1}{c}{\bf A5} & \multicolumn{1}{c}{\bf A6} & \multicolumn{1}{c}{\bf A7}  \\ 
			\hline
			\cite{yarinezhad2018reducing} & {\rotatebox[origin=c]{-90}{{Flat}}} & A single mobile sink moves randomly within a sensing field to periodically send a notification message to its one-hop neighbouring nodes. Information about the sink such as its latest position will be updated in order to transmit the sensed data to the sink efficiently.  & {\rotatebox[origin=c]{-90}{{Random}}}  & {\rotatebox[origin=c]{-90}{{NA}}}  & Transferring data within the shortest possible time, which improves the network lifetime. & Random mobility causes high latency between visiting times. \\ 
			\cite{chatzigiannakis2008efficient} & {\rotatebox[origin=c]{-90}{{Flat}}} & Method 1: A single mobile sink moves chaotically towards all directions in an angle of [-pi, pi] radians.  & {\rotatebox[origin=c]{-90}{{Random}}}  & {\rotatebox[origin=c]{-90}{{Constant}}}  &  It requires no network knowledge, guarantees visiting all sensors in the network and avoids the hot-spot problem. & Latency between consecutive visits to a sensor may be long, which causes a high packet-drop rate. \\ 
			\cite{chatzigiannakis2008efficient} & {\rotatebox[origin=c]{-90}{{Flat}}} & Method 2: A single mobile sink moves on a predefined trajectory to visit sensor nodes for data acquisition  & {\rotatebox[origin=c]{-90}{{Predicted}}} & {\rotatebox[origin=c]{-90}{{Constant}}} & The mobile sink moves closer to sensor nodes, which consequently reduces energy consumption. & The hot-spot problem still occurs at the nodes close to the fixed trajectories of the mobile sink. \\ 
			\cite{basagni2008controlled} & {\rotatebox[origin=c]{-90}{{Flat}}} & Method 1: The optimal sink trajectories can be found by using the mixed integer linear programming (MILP) formulation. & {\rotatebox[origin=c]{-90}{{Controlled}}} & {\rotatebox[origin=c]{-90}{{NA}}} & The optimal routes for the mobile sink and the sojourn times at stopping points can be obtained in order to enhance the network lifetime. & There is unbalanced energy consumption among nodes in the network. And nodes in regions where the mobile sink does not pass through may deplete their power faster than others in the network. \\ 
			\cite{basagni2008controlled} & {\rotatebox[origin=c]{-90}{{Flat}}} & Method 2: The next places selected for a mobile sink to visit are surrounded by high residual energy sensor nodes. & {\rotatebox[origin=c]{-90}{{Controlled}}} & {\rotatebox[origin=c]{-90}{{NA}}} & The balanced energy consumption among nodes in the network is carefully considered and hence this scheme can prolong the network lifetime. & The greedy maximum residual energy (GMRE) scheme does not work properly if there is lack of global knowledge of key network parameters. \\ 
			\cite{sharma2017rendezvous} & {\rotatebox[origin=c]{-90}{{Flat}}} & Method 1: Sensed data is transmitted to the closet backbone-tree node and then these backbone-tree nodes forward the data to a sink.  & {\rotatebox[origin=c]{-90}{{Random}}}  & {\rotatebox[origin=c]{-90}{{Adaptive}}}  & It can obtain high performance in terms of end-to-end
			latency and delivery ratio & Overloaded buffer can happen at a node if it receives data from many source nodes but the mobile sink does not visit it in time. \\ 
			\cite{sharma2017rendezvous} & {\rotatebox[origin=c]{-90}{{Flat}}} & Method 2: By retrieving location information of the mobile sink through the nearest backbone-tree nodes, a source node transmits data directly to the sink by a multi hop manner. & {\rotatebox[origin=c]{-90}{{Random}}}  & {\rotatebox[origin=c]{-90}{{Adaptive}}} & The network lifetime can be prolonged due to low energy consumption in the network. & Multi-hop communication protocol may lead to high latency in data transmission.  \\ 
			\cite{izadi2016alternative} & {\rotatebox[origin=c]{-90}{{Cluster}}} & A car based mobile sink moves around a network and collects sensed data directly from cluster head nodes.  & {\rotatebox[origin=c]{-90}{{Predicted}}}  & {\rotatebox[origin=c]{-90}{{Adaptive}}} & Most of data can be gathered with low latency. & Since the sink traverses on a fixed trajectory, the sensor nodes close to the sink may consume their power faster than the others. \\ 
			\cite{zahra2018integrated}. & {\rotatebox[origin=c]{-90}{{Flat}}} & Two metaheuristic approaches including tabu search and simulated annealing are employed to find the best travelling path for the sink.  & {\rotatebox[origin=c]{-90}{{Controlled}}}   & {\rotatebox[origin=c]{-90}{{Constant}}} & Both data collection and energy efficiency in WSNs are significantly increased. & Overloaded buffer and high latency are constraints of the proposed technique. \\ 
			\cite{kaswan2016routing} & {\rotatebox[origin=c]{-90}{{Flat}}} & A single mobile sink only travels to areas of interest by visiting some predefined RPs to collect data from local sensor nodes. & {\rotatebox[origin=c]{-90}{{Controlled}}} & {\rotatebox[origin=c]{-90}{{Constant}}} & RPs are selected from the heavily used sensor nodes, which reduces multi-hop data forwarding routes from other nodes to the sink. & It is complicated to select RPs in a large-scale network with many high load nodes. \\ 
			\multicolumn{7}{l}{\it A1: References; A2: Network structure; A3: Techniques.}  \\ 
			\multicolumn{7}{l}{\it A4: Mobility pattern; A5: Sink speed; A6: Advantages; A7: Disadvantages.} \\ 
			\hline
		\end{tabular}
	\end{center}
	\label{table3}
\end{table*}

\subsection{One Sink - Random Mobility}
\begin{figure}[tb]
	\centering
	\includegraphics[width=0.5\textwidth]{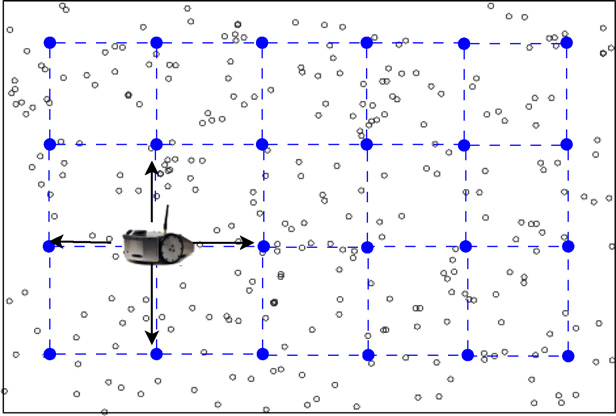}
	\caption{Possible sink movements on an overlay graph \cite{chatzigiannakis2008efficient} }\label{Figonesink_random}
\end{figure}

In the first mobility strategy, which is simple but unpredictable, similar to natural movements of many entities, the MS moves on a random path, regarding its position and speed. For instance, the authors in \cite{chatzigiannakis2008efficient,chatzigiannakis2006sink} proposed an efficient and robust approach to exploit a MS in data delivery in MWSN. It was proposed that the sink can move as the simple random walk, biased random walk or walks on spanning subgraphs. More specifically, the mobility function relied on a transition graph was proposed, where the network is partitioned in equally-sized areas and the center of each area is set as a vertex of the graph. An overlap graph is demonstrated in Fig. \ref{Figonesink_random}, where the sink is initialized on or near one of the vertices of the graph. Given a constant and predefined speed, the next stop of the sink is uniformly randomly selected by one of neighbors of the current position. Though it is dependent on data collection strategies, the proposed scheme can save up to 30 percent of energy consumption in the network.

One of challenges in WSNs given a MS is the dynamic network topology. That is, in order to efficiently deliver data from the sensor nodes to the sink, the latest sink location is required to update to the sensor nodes, which causes high energy consumption and data latency in the network. To better address this issue as compared with techniques such as the periodic flooding \cite{coffin2000declarative,ye2005gradient,basagni2008controlled}, Yu \textit{et al.} in \cite{yu2010simple} developed a new scheme based on overhearing feature of wireless transmission to propagate the location information of the sink to the other nodes, which leads to reduce energy consumption and increase data delivery rate. Likewise, Yarinezhad \textit{et al.} in \cite{yarinezhad2018reducing} proposed a novel routing method that relies on a virtual grid infrastructure. Similar to the work \cite{chatzigiannakis2008efficient}, the authors proposed to discretize the network into equal regions; and the closest nodes to intersection of four regions, which are considered as the nodes of the virtual infrastructure, are assumed to record the latest position of the sink. The proposed paradigm guarantees that all the virtual infrastructure nodes can be properly distributed in the network, and other sensor nodes can be known the last sink location with a minimum number of hops from the nearest node in the virtual infrastructure. The proposed algorithm can be applicable for any network size.

In contrast to the equally-sized regions in the network presented in the aforementioned works, Sharma \textit{et al.} in \cite{sharma2017rendezvous} proposed a rendezvous-based routing protocol (RRP) to improve energy efficiency and data end-to-end latency in mobile wireless sensor networks (MWSNs). They defined a rendezvous region, also called a virtual cross or backbone-tree area, in the middle of the network. And data from the sensor nodes is transmitted to the MS via the backbone-tree region. Though the MS is randomly moving in the network, its location information is continuously recorded by the backbone-tree nodes so that the sink is always able to receive the sensor readings. For instance, when the sink moves to a new locations, it send a message of its locations to one of its neighbour nodes, which is selected based on a location factor, computed as follows,
\begin{equation}
node={\operatorname{\textbf{argmax}}}\; \frac{Er_i}{Er_{max}\sqrt{(x_d-x_i)^2+(y_d-y_i)^2}},
\end{equation}
where $Er_i$ is the residual energy of node $i$ at the location $(x_i,y_i)$, and the node $i$ is one of neighbour nodes of the sink. $Er_{max}$ is the maximum residual energy in the all possible nodes $i$. $(x_d,y_d)$ is the destination location. The selected node then forwards the sink location to one of its neighbors, following the same procedure. The process repeats until the location information reaches to one of the backbone-tree nodes. Given random movements of the MS, the proposed RRP can attain diverse objectives in data wireless transmission such as high delivery ratio and no hot-spot problem in WSN. Nonetheless, it still reveals some limitations in improving the network lifetime, where energy dissipation of the backbone-tree nodes may increase if the source nodes are far from the sink and vice versa. Moreover, the high end-to-end latency and buffer overflow problems may result in loss of information in the event areas.   	

To compare performances of the mobility models in the mobile ad hoc networks, the authors in \cite{rathore2012comparative} provided a comprehensive analysis with demonstrated results on two categories of entity and group mobility models, given different routing schemes. Overall, the group mobility models (e.g. pursue and column) outperform the entity mobility models (e.g. random walk, restricted random walk and random direction). It is noted that the mobility models presented in \cite{rathore2012comparative} demonstrate ability to address the hot-spot problem in WNSs, which eventually elongates the network lifetime. Nevertheless, their disadvantages comprise a long latency period and high ratio of dropped packets.

In fact, the random or unpredictable mobility strategy for a MS is widely used in the realistic applications of monitoring environmental parameters such as temperature, humidity, light, wind and so on, where behaviours of movements are natural. However, due to its random pattern, no one can guarantee that the sink is able to reach all the sensor nodes in the sensing field given its time and energy constraints. In the following section, the sink moving on a predefined path will be discussed. 

\subsection{One Sink - Fixed/Predictable Mobility}
In contrast to the random mobility, the fixed/predictable mobility is expected to drive the MS on a predefined path. In other words, the sensor nodes may know the expected visiting time of the sink and then optimize their sensing and data transmitting tasks, which leads to minimum protocol overhead. This mobility pattern can provide boundedness of the transmission delay and elongate the network lifetime. However, due to the fixed path, in some occasions the sensors close to the sink may not have data while the others have to transmit their data through multi-hope routing \cite{izadi2016alternative}.

\begin{figure}[tb]
	\centering
	\includegraphics[width=0.5\textwidth]{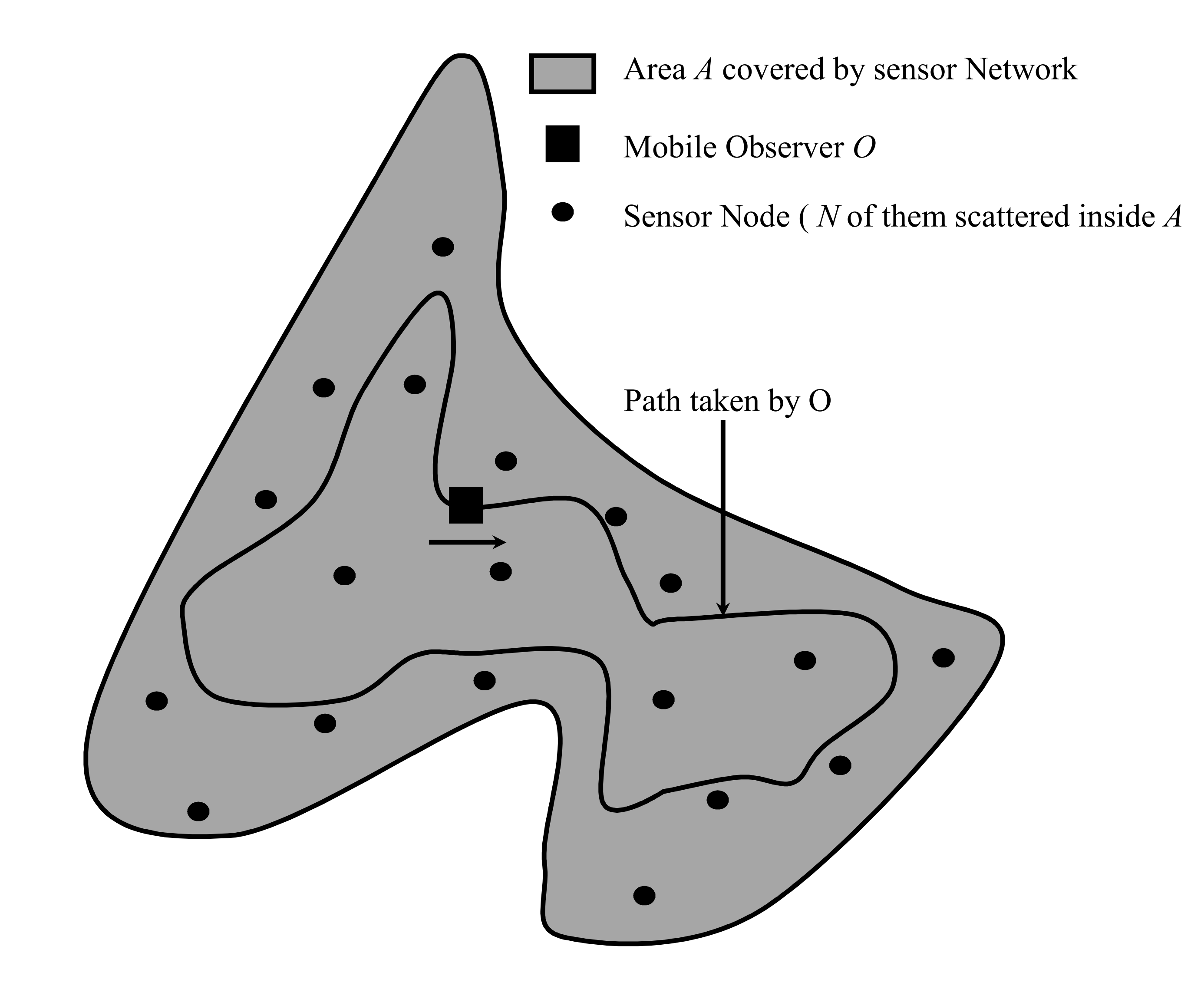}
	\caption{A possible path for the MS in a WSN \cite{chakrabarti2006communication}}\label{Figonesink_fixed1}
\end{figure}

In the works \cite{chakrabarti2006communication,chakrabarti2003using} Chakrabarti \textit{et al.} assumed that the MS always traverses on a pre-specified path as illustrated in Fig. \ref{Figonesink_fixed1} and proposed the queuing model, which is exploited with other system parameters to ascertain adequate data collection in areas of interest with minimum power. That is, a simple MS driven routing protocol was proposed to wake up the sensor node for data transmission when the sink is nearby.

On other works such as \cite{luo2005joint,wang2008extending,holla2017data}, the authors employed the multi-hop routing protocol and mobility of the sink to enhance the network longevity. For instance, a joint routing and mobility approach was proposed for the data collection protocol, which analytically demonstrates load balance in the network. More importantly, since a multi-hop routing strategy is in use, the data latency is not influenced and the MS location known by all the sensor nodes is not necessary. The joint mobility and routing algorithm can improve the network lifetime up to a factor of four \cite{wang2008extending} or 500 \% \cite{luo2005joint}.

In this mobility pattern, it seems that the MS collecting data from the sensor nodes through rendezvous points (RPs) is preferred. For instance, in \cite{wen2017energy}, a simple but efficient data collection scheme was proposed by electing the important sensor nodes as RPs, which considers their locations as compared with the others. The traveling path for the sink is constructed based on the RPs while unselected sensors send their readings to the nearest RP. In the cases where the sensor nodes gather data unevenly, Kumar \textit{et al.} in \cite{kumar2018aco} proposed an algorithm based on the ant colony optimization to find a near optimal set of RPs to building a data collecting path for the MS. The proposed method allows a WSN to near-optimally maximize the network lifetime and minimize the delay in gathering sensory data from the nodes. One of advantages of the proposed approach as compared to others is that the RPs are re-elected after every iteration, which leads to balancing energy consumption in the whole network.

\begin{figure}[tb]
	\centering
	\includegraphics[width=0.45\textwidth]{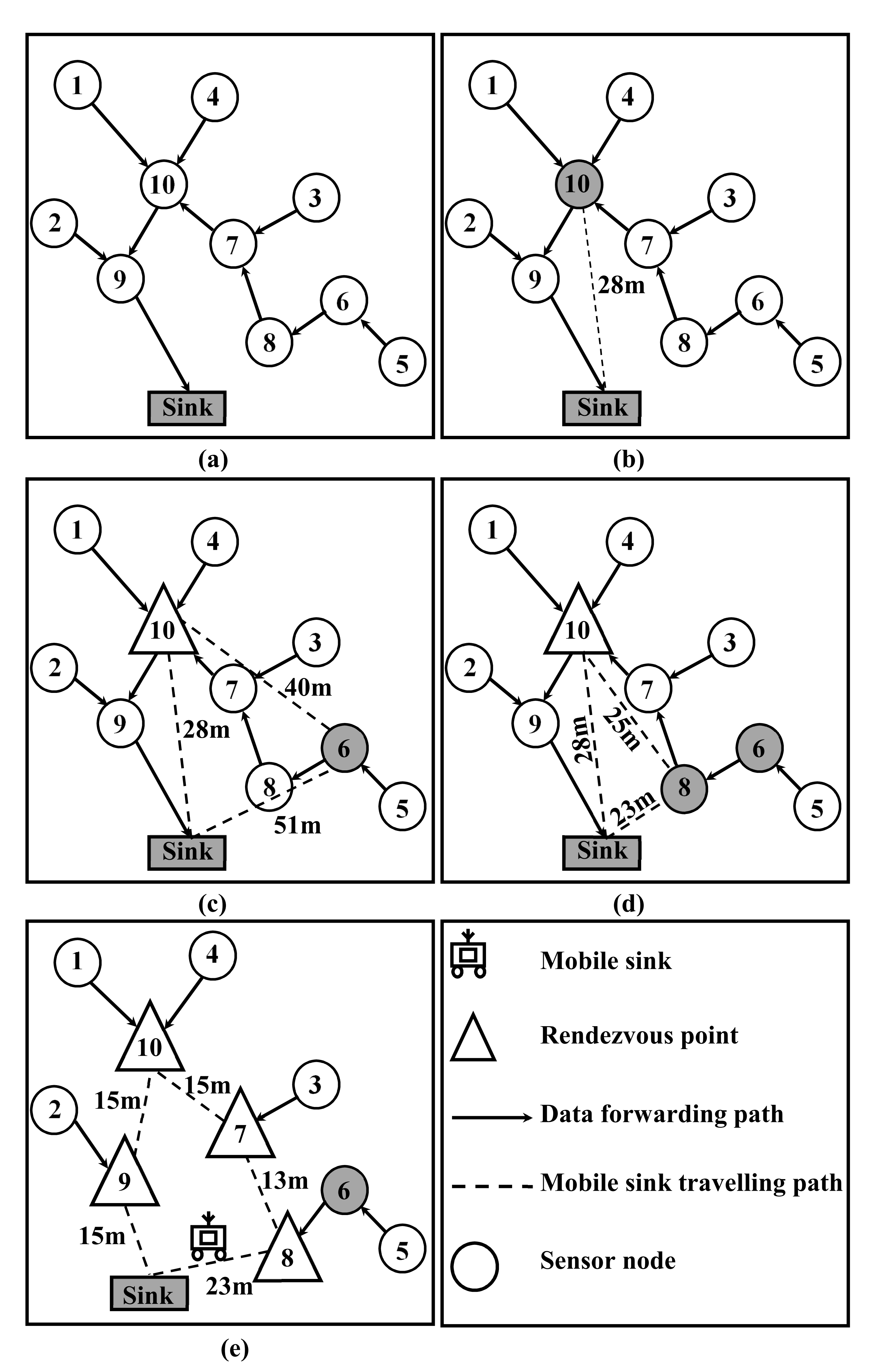}
	\caption{An example of the weighted rendezvous planning algorithm in a WSN with 10 sensor nodes \cite{salarian2014energy}}\label{Figonesink_fixed3}
\end{figure}

Similarly, given a latency bound, the MS is expected to visit all the RPs in the network, as presented in \cite{salarian2014energy}. Although proving that finding an optimal set of RPs is a NP-hard problem, the authors proposed a heuristic weighted planning algorithm to find its near-optimal solution. The weight for each sensor node is computed based on its hop distance and number of data packets intended to transmit to the nearest RP. the weight of the sensor node $i$ is computed as follows,
\begin{equation}
w_i=(c(i,T_{r_j})+1)h(i,R),
\end{equation}
where $c(i,T_{r_j})$ is the number of data packets forwarded by the neighbor nodes of the sensor $i$ to the node $i$. It is noted that the sensor node $i$ transmits all the data it received from the neighbors and its own to the nearest RP $r_j$ through the routing tree $T_{r_j}$ with the root at $r_j$. $R$ is the set of all the RPs in the network, and $r_j\in R$. $h(i,R)$ is the hop distance from the sensor node $i$ to the traveling path for the sink, built on the set of RPs $R$, and computed by
\begin{equation}
h(i,R)=\lbrace h_{i,r_j}\vert\forall r_k\in R, \; h_{i,r_j}\leq h_{i,r_k}\rbrace.
\end{equation}
One example of implementing the weighted rendezvous planning algorithm in a WSN with 10 sensor nodes is demonstrated in Fig. \ref{Figonesink_fixed3}. Note that the proposed method is able to increase the network longevity by 44 \% and decrease the network energy consumption by 22 \% as compared with other techniques.

Another work designing the MS traveling path based on RPs is \cite{kaswan2018multi}, where a new evolutionary algorithm derived from the multi-objective particle swarm optimization (MOPSO) was proposed to find a near-optimal solution for the NP-hard problem selecting RPs. The proposed method computes the efficient particle encoding scheme by exploiting Pareto dominance, which provide each particle with both global and local best guides. In other words, each sensor node gathers information of its neighbors and conveys it to the sink, where those information is utilized to select RPs. Nevertheless, since the particle elements are randomly generated, the RPs may be arbitrarily elected, which results in the non-uniform RPs distribution and biased energy consumption in the network.

The traversing trajectory for the MS is constructed not only by RPs but also by clustering. For instance, in \cite{wang2017particle}, Wang \textit{et al.} proposed an efficient routing method based on particle swarm optimization and mobility of the sink. In equivalent words, to prolong the network longevity, the network is divided into regions, where a cluster head (CH) is elected based on its own residual energy and distance to the gravity center. Then the MS moves on a pre-defined trajectory to collect data from the CHs. The proposed algorithm also developed different data packet formats to define when to send data to the CH and when to send data to the sink, which ultimately ameliorate the transmission latency. In a dynamic fashion, Holla \textit{et al.} in \cite{holla2017data} developed an approach to extend lifetime of the network by the use of harmony search. In the proposed method, a CH for each clustering area is selected based on its location so that distribution of the CHs in the whole network is even. More importantly, difference from the work \cite{wang2017particle} is that the CH in each cluster will be re-elected at every iteration if there is any node in that cluster died, which leads to balance of energy consumption among the rest.

Though denoted by different terms, RP and CH play a quite similar role in a MWSN. In \cite{kaswan2017energy}, the authors developed two new algorithms to balance between the traveling path length of the sink and the data transmission path lengths of the sensor nodes. To ensure coverage of the entire network, RPs are selected by the use of the k-means clustering and weight function, computed from the deployed sensor nodes. More particularly, other factors such as minimizing overall hop counts, average hop distance and buffer overflow are also considered in the routing protocol. Nonetheless, though both the network lifetime and ratio of dropped packets are enhanced, the proposed algorithm has not taken the different number of packets at each sensor node into account, which may cause sojourn time of the sink at each RP to be diverse in different regions.
	
On the other hands, some irregular ways to define a traversing trajectory for the MS have also been presented in literature. For example, a reactive data forwarding technique \cite{giannakos2009message} enables the sink to send request messages to a sensor node located in vicinity of its future position. That node will gather all the data from its vicinity and transmit them to the sink when it is nearby to avoid the delay. In a similar manner, Xu \textit{et al.} in \cite{5982160} considered several nodes along the sink path as gateways or relay nodes. Only data packets with the maximum quality are collected by the gateways and then transmitted to the sink when it passes by. Furthermore, the work \cite{chang2012ebdc} discusses an energy-balanced data collection algorithm to prolong the network operational time by balancing the data-relaying workloads in the network. The sensing field is first divided into similar circular tracks, then the sink is driven along different tracks with pre-specified sweep repetitions. The sink trajectory can also be formed by the use of Hilbert space filling curve as presented in \cite{ghafoor2014efficient}. The Hilbert curve order is first computed relied on the network size, which is then employed to dynamically construct a path for the sink. With a high curve order of the MS trajectory, the proposed approach increases the network coverage, the packet delivery ratio as well as reduces the energy consumption. Nonetheless, the number of visiting points of the MS increases with the curve order; therefore, the single MS may not visit all points in a predefined time, which causes a high dropped packet ratio and may lose critical information from some event areas.

A single MS in a WSN is efficiently utilized not only for data collection but also sensor localization. In many WSN applications, geolocation data is required \cite{Nguyen2018a, Nguyen2017b, Nguyen2012a, Nguyen2014c, Nguyen2013}. Thus, a MS is expected to play as an anchor node moving around the network to collect information of sensor locations \cite{han2017disaster}. To define a path for the anchor node, a localization algorithm was designed by the use of trilateration, where a location of an unknown sensor node is formed by beacon packets obtained at three points on vertices of a regular triangle. Addressing the localization problem in a WSN by employing a MS can attain highly accurate results with reduced energy consumption.

The sink mobility on a fixed or predictable path is widely utilized in realistic applications. For instance, in some military systems, sensor nodes can be deployed along highways, bridges or rivers to track targets in a battlefield.

\subsection{One Sink - Controlled Mobility}
In controlled mobility, the trajectories, velocities and directions of a MS are dynamically and optimally computed based on the network constraints including delay, throughput, power consumption, hop counts and so on. In other words, the source-to-sink routing paths are optimized, given a change of parameters of interest in network state, which leads to significantly reduced energy consumption. Though optimizations in the controlled mobility are complicated, this pattern is superior to the random and predicted mobility models.

\begin{figure*}[tb]
	\centering
	\includegraphics[width=0.8\textwidth]{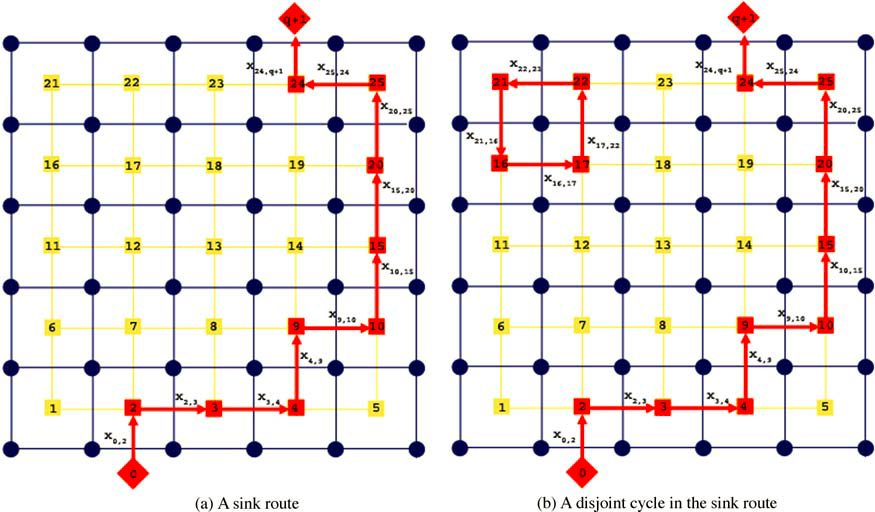}
	\caption{Optimal sink trajectories obtained by the network constraints \cite{basagni2008controlled}}\label{Figonesink_control1}
\end{figure*}

In the first instance, multiple network constraints are considered in optimizing the sink mobility pattern in the works \cite{chen2017lifetime,zahra2018integrated}. More specifically, Zahra \textit{et al.} in \cite{zahra2018integrated} proposed an optimality criterion to define the best traveling path for the sink, which aims to optimize three parameters including minimum traversing path for the sink, minimum distances from the sink to sensor nodes with the lowest error rate and maximum signal quality. The optimization problem was then addressed by a metaheuristic approach such as tabu search or simulated annealing, which results in energy efficiency. Likewise, the authors in \cite{chen2017lifetime} developed an optimization problem for the network longevity, which is constrained by data transmission, data transmission time, node coverage, energy consumption, and grid selection. Note that in this proposition, environment is discretized into grids, where the MS can gather data from grid centers. By the use of maximum capacity path \cite{wang2014network} and genetic algorithms, the proposed optimization can be solved, where sojourn grid centers and sojourn time for the MS can be attained, which leads to the network elongation.\\
In another context, the MS is proposed to collect data from RPs that are selected in conjunction with the network parameter constraints. For instance, both the works \cite{kaswan2016routing} and \cite{xing2008rendezvousc} proposed that the sink should be able to gather data with a delay bound. To this end, in \cite{kaswan2016routing}, RPs are selected from the heavily used sensor nodes, which decreases multi-hop data forwarding routes from the nodes to the sink. The selection enables the network to balance routing load when data is collected. Nonetheless, RPs can be on an approximate Steiner minimum tree of source nodes \cite{xing2008rendezvousc}, which leads to a shorter travelling path for the sink while collecting sensor readings. Moreover, an idea of constructing clusters in the network was implemented in \cite{zhu2015tree,almi2010energy}, where data in a cluster is transmitted to a cluster representative. While \cite{almi2010energy} arranged sensor nodes into balanced-size groups through an integer linear programming optimization problem, \cite{zhu2015tree} built a cluster tree that jointly takes densities of local nodes, distances from nodes to the sink and residual energies of nodes into account. More importantly, subrendezvous points are also appointed given nodes' hop counts and data amounts to further balance energy dissipation in the network.
It has been realized that the sensor nodes closer to the sink consume more energy than the others due to multi-hop data transmission. Therefore, some research works designed a specific trajectory for the MS, where the sink is driven to approach the sensor nodes that have higher residual energy \cite{basagni2008controlled,mudigonda2011mobility,bi2007hums}, which eventually prevents the network from a hotspot problem. For instance, in \cite{bi2007hums}, in each iteration the sink autonomously computes its next stop relied on energy information at each node that the sensor sent together with its data packets in the previous iteration. Similarly, \cite{mudigonda2011mobility} proposed to exploit maximum residual energy in each region of interest to decide where the MS should move to. Moreover, to improve the network lifetime, \cite{basagni2008controlled} mathematically defines a mixed integer linear programming analytical model for the sink movements, as follows,
\begin{subequations}
\label{onesink_control_equ10}
\begin{alignat}{2}
&\textbf{max}        &\qquad& \sum_{k\in S} t_k \\
&\textit{subject to} &      & \sum_{k\in S}c_{ik}t_k+\sum_{k\in S}f_{ik}y_k\leq e_0 \;\; (i\in N)\\
&                  &      & t_{min}y_k\leq t_k\leq My_k \;\; (k\in S) \\
& & &\sum_{k\in S}x_{0k}=1 \\
& & & \sum_{k\in S}x_{k,q+1}=1 \\
& & &\sum_{\substack{j\in S\cup\{0\}\\(j,k)\in O\cup A}}x_{jk}=\sum_{\substack{j\in S\cup\{q+1\}\\(k,j)\in A\cup D}}x_{kj} \;\; (k\in S) \\
& & &\sum_{\substack{j\in S\cup\{0\}\\(j,k)\in O\cup A}}x_{jk}=y_k \;\; (k\in S) \\
& & & u_j-u_k+qx_{jk}\leq q-1 \;\; ((j,k)\in A)\\
& & & t_k,u_k\geq 0 \;\; (k\in S) \\
& & & y_k\in\{0,1\} \;\; (k\in S) \\
& & & x_{j,k}\in\{0,1\} \;\; ((j,k)\in X),
\end{alignat}
\end{subequations}
where $S$ is the set of possible MS locations, and $N$ is the number of the sensor nodes. $e_0$ is initial node energy while $f_{i,k}$ and $c_{i,k}$ are energy dissipations at the node $i\in N$ for setting up routes and receiving and transmitting data packets when the sink approaches to the location $k\in S$, respectively. $t_{min}$ and $t_k$ denote sojourn time and minimum sojourn time when the sink is at the location $k\in S$. $A=\{(j,k)\in S\times S: j\neq k, d_{jk}\leq d_{max}$, where $d_{jk}$ and $d_{max}$ are the distance and maximum distance between two any locations $j,k\in S$. While $O$ and $D$ define sets of distances from origin to locations in $S$ and from locations in $S$ to the destination, respectively, $X$ is union of $A$, $O$ and $D$. $y_k=1$ if the sink is at location $k$, and $x_{jk}=1$ if the sink traverses from $j$ to $k$. $u_k$ is randomly set when the sink is required to travel on a unique path. It is note that the optimization problem (\ref{onesink_control_equ1}) is resolved by the proposed greedy maximum residual energy heuristic algorithm, which aims to drive the sink toward the area with highest residual energy. Examples of the optimal MS paths obtained by the optimization criterion in (\ref{onesink_control_equ1}) are presented Fig. \ref{Figonesink_control1}.

In some applications such as habitat monitoring, where the MS is expected to minimize disturbance to the targeted animal species, MWSNs are designed to tolerate delay of data delivery. In other words, sensor nodes are expected to sense information and temporarily store itself. They only transmit those data to the sink when it approaches them at convenient time to prevent the system from disturbing targets. To this end, \cite{achour2017improved} proposed a new linear programming model to present network energy dissipation, where the optimization problem is constrained by sojourn time at each sensor node, which defines how much time the sink is allowed to collect data. Furthermore, beyond constraint of data collection delay, the optimality criterion can be constrained by energy or flow conversation \cite{yun2010maximizing} to maximize energy efficiency in the network.

It has been shown that a good practical routing protocol can be able to elongate the network lifetime. For instance, \cite{mitra2018proactive} proposed a method that considers delayed information delivery from nodes to the sink in conjunction with a hierarchical routing protocol based on a virtual grid, which is utilized to find an optimal traveling path for the MS. The premise behind the approach is taking both hop counts and data generation rates obtained from priori upstream information into account. In equivalent words, the sensing field is divided into a virtual grid, where a RP is elected for each grid to gather all data from that grid area. Due to delay requirements of the sink, only a small group of RPs are used to form the sink path while the others are required to send data to their nearest counterparts. This hierarchy of data collection enables the system to enhance throughput as well as data traffic. The linear programming optimization problem given a constraint of a routing protocol can be specified as follows \cite{luo2006mobiroute},
\begin{subequations}
\label{onesink_control_equ1}
\begin{alignat}{2}
&\textbf{max}        &\qquad& \sum_{n} t_n \\
&\textit{subject to} &      & \sum_{n} t_nP_n\leq E,
\end{alignat}
\end{subequations}
where $P_n$ is the power dissipation of the node $n$ when the sink visits it during the sojourn time $t_n$, while $E$ is its initial energy.

The mobility in a MWSN has attracted much attention in the research community due to its capability to be efficiently implemented in various applications \cite{natalizio2013controlled, goldenberg2004towards,rao2004purposeful,somasundara2006controllably, bisnik2007stochastic,butler2004controlling,cortes2004coverage, 1636223, luo2005joint, basu2004network, zhao2005controlling}. Nevertheless, there are some issues that needs to be understood before implementing this strategy. For instance, additional hardware is required in installation \cite{costa2012use}, and there is possibility of disconnection in the network due to the sink movements, which may lead to dropped packets during data transmission.

\section{Multiple Sink Mobility based Efficient Energy}
\label{sec_5}
It has been learned that employing a MS in a MWSN considerably benefits not only high data delivery ratio but also low end-to-end latency. Nonetheless, a single sink traversing in a large-scale network is hardly able to gather all data from sensor nodes under time constraints. That leads to proposition of utilizing multiple sinks simultaneously navigating through the network in a given time for data collection. The following subsections will survey how to cooperatively drive many sinks in a sensing field so that energy consumption at the sensor nodes is minimized while all sensor readings are effectively gathered. Brief information of the existing techniques is summarized in Table \ref{table4}.

\begin{table*}
	\caption{TECHNIQUES FOR IMPROVING NETWORK LIFETIME BASED ON MULTIPLE SINK MOBILITY}
	\begin{center}
		\begin{tabular}	{m{0.25cm}m{2cm}m{0.25cm}m{0.25cm}m{0.25cm}m{2cm}m{0.5cm}m{2cm}m{4.8cm}}
	\hline
	\multicolumn{1}{c}{\bf A1} & \multicolumn{1}{c}{\bf A2} & \multicolumn{1}{c}{\bf A3} & \multicolumn{1}{c}{\bf A4} & \multicolumn{1}{c}{\bf A5} & \multicolumn{1}{c}{\bf A6 (nodes/m2)} & \multicolumn{1}{c}{\bf A7} & \multicolumn{1}{c}{\bf A8 (m/s)} & \multicolumn{1}{c}{\bf A9} \\ 
	\hline
	\multicolumn{1}{c}{\cite{jain2006exploiting}} & \multicolumn{1}{c}{Flat} & \multicolumn{1}{c}{No} & \multicolumn{1}{c}{Yes} & \multicolumn{1}{c}{No} & 0.025 & \multicolumn{1}{c}{No} & \multicolumn{1}{c}{Constant (10)} & Avoiding the hot-spot problem completely. \\ 
	\multicolumn{1}{c}{\cite{banerjee2010increasing}} & \multicolumn{1}{c}{Cluster based} & \multicolumn{1}{c}{Yes} & \multicolumn{1}{c}{No} & \multicolumn{1}{c}{No} & 0.0033 & \multicolumn{1}{c}{No} & \multicolumn{1}{c}{NA} & Avoiding the hot-spot problem completely with randomly moving cluster head nodes. \\ 
	\multicolumn{1}{c}{\cite{xie2017energy}} & \multicolumn{1}{c}{Flat} & \multicolumn{1}{c}{No} & \multicolumn{1}{c}{Yes} & \multicolumn{1}{c}{Yes} & 0.017 & \multicolumn{1}{c}{Yes} & \multicolumn{1}{c}{Constant} & Avoiding the hot-spot problem completely. \\ 
	\multicolumn{1}{c}{\cite{gao2011efficient}} & \multicolumn{1}{c}{Cluster based} & \multicolumn{1}{c}{No} & \multicolumn{1}{c}{Yes} & \multicolumn{1}{c}{No} & 0.0005-0.00083  & \multicolumn{1}{c}{Yes} & \multicolumn{1}{c}{Constant (5)} & The hot-spot problem may happen at nodes close to trajectories of the sinks.  \\ 
	\multicolumn{1}{c}{\cite{kinalis2007scalable}} & \multicolumn{1}{c}{Cluster based} & \multicolumn{1}{c}{No} & \multicolumn{1}{c}{NA} & \multicolumn{1}{c}{No} & 0.0075  & \multicolumn{1}{c}{Yes} & \multicolumn{1}{c}{Constant (4)} & Avoiding the hot-spot problem completely. \\ 
	\multicolumn{1}{c}{\cite{lee2018active}} & \multicolumn{1}{c}{Cluster based} & \multicolumn{1}{c}{No} & \multicolumn{1}{c}{No} & \multicolumn{1}{c}{No} &  0.00044 & \multicolumn{1}{c}{Yes} & \multicolumn{1}{c}{Constant (10)} & The hot-spot problem may happen at nodes in the local data area. \\ 
	\multicolumn{1}{c}{\cite{ye2002two}} & \multicolumn{1}{c}{Cluster based} & \multicolumn{1}{c}{Yes} & \multicolumn{1}{c}{Yes} & \multicolumn{1}{c}{Yes} & 0.000025 & \multicolumn{1}{c}{No} & \multicolumn{1}{c}{Constant (10)} & A separate grid for each source. \\ 
	\multicolumn{1}{c}{\cite{han2013low}} & \multicolumn{1}{c}{Flat} & \multicolumn{1}{c}{Yes} & \multicolumn{1}{c}{No} & \multicolumn{1}{c}{Yes} & 0.0015-0.01 & \multicolumn{1}{c}{Yes} & \multicolumn{1}{c}{Constant (1)} & Avoiding the hots-pot problem completely. \\ 
	\multicolumn{1}{c}{\cite{marta2009improved}} & \multicolumn{1}{c}{Cluster based} & \multicolumn{1}{c}{No} & \multicolumn{1}{c}{Yes} & \multicolumn{1}{c}{Yes} & 0.021 & \multicolumn{1}{c}{Yes} & \multicolumn{1}{c}{Constant} & The hot-spot issues are moderated with movements of the sinks along the hexagon perimeters. \\ 
	\multicolumn{1}{c}{\cite{marta2009improved}} & \multicolumn{1}{c}{Cluster based} & \multicolumn{1}{c}{Yes} & \multicolumn{1}{c}{Yes} & \multicolumn{1}{c}{Yes} & 0.021 & \multicolumn{1}{c}{Yes} & \multicolumn{1}{c}{Constant} & The hot-spot problem may happen when movements of the sinks are autonomous. \\ 
	\multicolumn{1}{c}{\cite{Bi2007}} & \multicolumn{1}{c}{Flat} & \multicolumn{1}{c}{Yes} & \multicolumn{1}{c}{Yes} & \multicolumn{1}{c}{Yes} & 0.0025 & \multicolumn{1}{c}{Yes} & \multicolumn{1}{c}{NA} & Avoiding the hot-spot problem completely. \\ 
	\multicolumn{1}{c}{\cite{Wang2014}} & \multicolumn{1}{c}{Flat} & \multicolumn{1}{c}{No} & \multicolumn{1}{c}{No} & \multicolumn{1}{c}{Yes} & 0.005-0.015 & \multicolumn{1}{c}{Yes} & \multicolumn{1}{c}{NA} & Burden of the hot-spot issues is relieved. \\ 
	\multicolumn{1}{c}{\cite{Abuarqoub2017}} & \multicolumn{1}{c}{Cluster based} & \multicolumn{1}{c}{Yes} & \multicolumn{1}{c}{Yes} & \multicolumn{1}{c}{Yes} & 0.0004 & \multicolumn{1}{c}{Yes} & \multicolumn{1}{c}{Constant (10)} & The hot-spot problem is  mitigated. \\ 
	\multicolumn{1}{c}{\cite{Xing2008}} & \multicolumn{1}{c}{Flat} & \multicolumn{1}{c}{Yes} & \multicolumn{1}{c}{Yes} & \multicolumn{1}{c}{Yes} & 0.0008 & \multicolumn{1}{c}{Yes} & \multicolumn{1}{c}{(0.1-0.5)} & Avoiding the hot-spot problem completely. \\ 
	\multicolumn{1}{c}{\cite{Liang2011}} & \multicolumn{1}{c}{Cluster based} & \multicolumn{1}{c}{Yes} & \multicolumn{1}{c}{No} & \multicolumn{1}{c}{Yes} & 0.0031-0.012 & \multicolumn{1}{c}{Yes} & \multicolumn{1}{c}{NA} & The balanced load among bottleneck sensor nodes is improved. \\ 
	\multicolumn{9}{l}{\it A1: References; A2: Network structure; A3: Sensor node location known; A4: Data aggregation; A5: Moving distance minimized;} \\ 
	\multicolumn{9}{l}{\it A6: Network density; A7: Energy aware routing; A8: Sink speed; A9: Hot-spot mitigation.} \\ 
	\hline
\end{tabular}
	\end{center}
\label{table4}
\end{table*}

\subsection{Multiple Sinks - Random Mobility}
Multiple MSs traversing randomly in the network without specific destinations are challenging issues in a MWSN. This section will investigate how the sinks are driven in a sensing field so that data is efficiently collected with minimum energy dissipation in the network.

\begin{figure}[tb]
	\centering
	\includegraphics[width=0.3\textwidth]{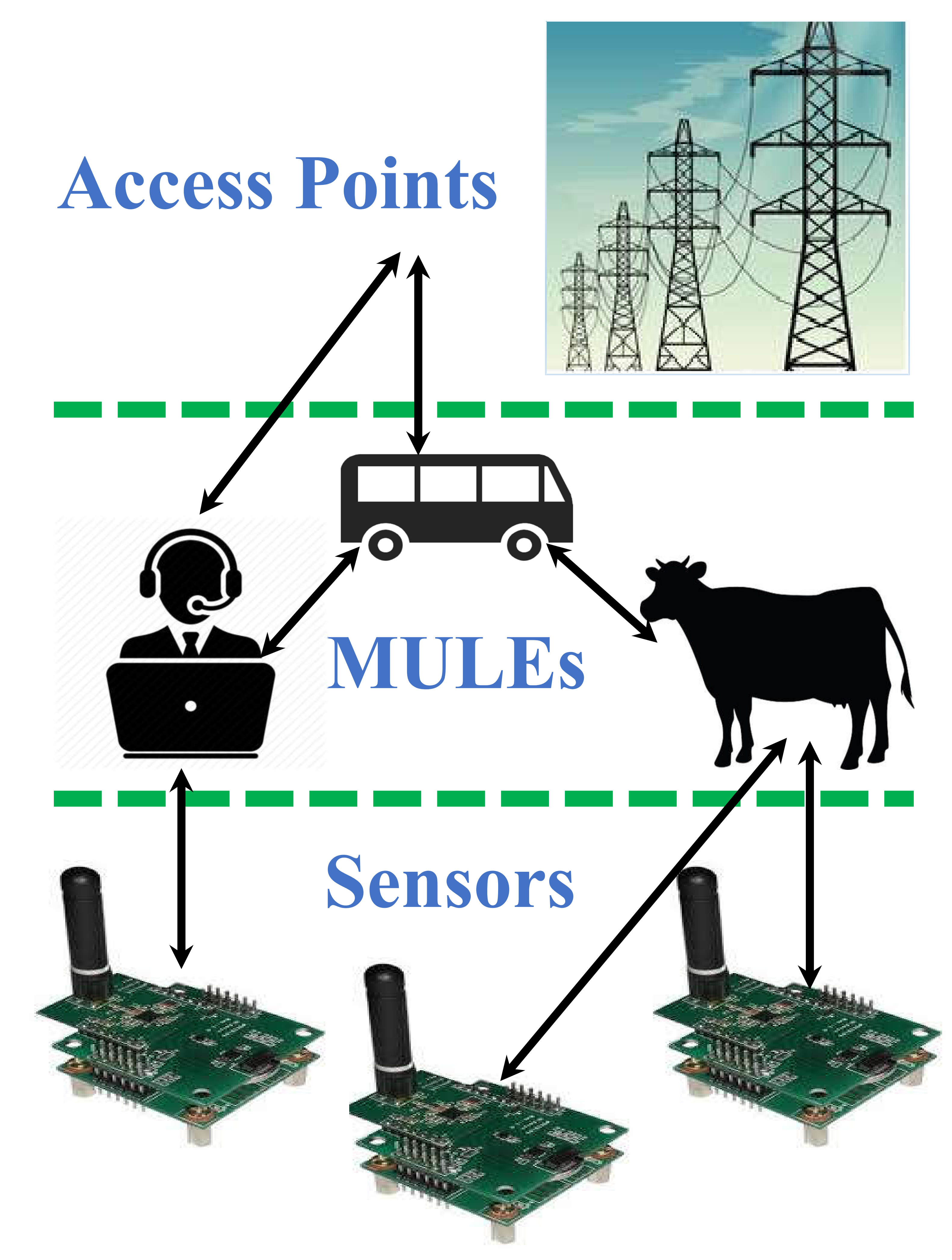}
	\caption{Multiple sinks in realistic applications \cite{jain2006exploiting}}\label{Figmultisink_random}
\end{figure}

In the context of network architecture, the MSs are not only mobile robotic platforms but also nature-based platforms in various realistic applications. For instance, in \cite{jain2006exploiting}, Jain \textit{et al.} discussed a three-tier layer architecture, employing mobile ubiquitous LAN extensions (MULEs) to effectively gather sensor readings in a network. In this design, MULEs, as demonstrated in Fig. \ref{Figmultisink_random} which can be animals in habitat or environmental monitoring applications or vehicles in air pollution monitoring study, are considered as the MSs. MULEs are expected to pass by sensor nodes so that data can be transferred to them in a short communication. Moreover, when they travel close by a higher tier layer e.g. access points, the sensor readings are unloaded from them. In the proposed paradigm, the random mobility patterns of MULEs can be random waypoint, random walk (i.e. humans, animals), deterministic travel (i.e. vehicles) or Poisson arrival. Given its short-range communication model, MULE design can effectively address hotspot problems and avoid communication collisions in the network. It is also able to enhance quality of service with spatial reuse and scalability.

Likewise, \cite{tong2003sensor} developed the sensor network with mobile agents (SENMA) scheme to extend network lifetime in a MWSN. In SENMA, it proposes that the MSs play a role of receiving terminals, which allows sensor nodes not to need to receive packet transmission request from their neighbors as compare with those in a flat ad hoc paradigm. Furthermore, the proposed scheme exploits free space communications between nodes and sinks, which declines energy consumption from 4th power of distance in a flat ad hoc sensor network with ground-reflected rays to 2nd power of distance. It proved that the proposed architecture substantially reduces energy waste in a flat ad hoc network.

In addition to architecture, a routing protocol for a network with multiple MSs has also been attracted research attention. To extend the network lifetime, Kumar \textit{et al.} in \cite{kumar2016improving} proposed a hierarchical ring routing for a large-scale multiple sink network structure. The proposition can be effectively employed in multi-target tracking applications, and it proved that longevity of a network can be enhanced up to 200\%. More particularly, the work \cite{lee2018active} proposed an active data dissemination protocol for a network of MSs. The premise behind the proposition is that all sensor nodes know the moving direction and pattern of the sinks, which is informed by a sink leader. The network establishes grid based local data areas, where it is predicted that the sinks will go through. All the sensor nodes send their readings to those local data areas, which will be picked up by the sinks in one-hop communication while they are passing those localities. Given the short transmissions between the nodes and sinks, energy in the network can be substantially saved; nevertheless, the sinks are expected to move slowly so that their moving directions can be accurately estimated. More generally, the authors in \cite{yang2017practical} developed a new contact-aware expected transmission count routing metric that aims to improve data collection in a high-throughput and low-delay opportunistic MWSN, where the network connections may be intermittent and number of the MSs and their movements are typically arbitrary.

In contrast to \cite{lee2018active}, the work in \cite{ye2002two} designed a two-tier data dissemination structure, where the sensing field is divided into grid cells. Only sensor nodes at grid vertices, considered as dissemination nodes, are required to know moving directions of the sinks. When a sink is close by dissemination nodes, a data query from the sink will be forwarded to the sources through the dissemination nodes, which significantly reduces energy consumption and network overload. Han \textit{et al.} in \cite{han2013low} also considered a new routing topology for a MWSN, which is a minimum Wiener index spanning tree. Although finding that tree is NP-hard, the authors proposed to utilize the branch and bound and simulated annealing algorithms to approximately obtain solutions.

In order to investigate the use of the sink multiplicity in the random mobility pattern, the work \cite{kinalis2007scalable} considered three possible scenarios in a MWSN. In the first strategy, which requires many MSs to collect all available data, each sink movement follows a simple random walk. In the second scheme, it is proposed to utilize a less number of MSs as compare with the first strategy, but combine with static sinks. In both the scenarios, the schemes can greatly reduce latency and increase data delivery rate; nonetheless, there is no connection among the MSs, which may cause missed visits on some nodes. Therefore, the authors proposed to employ beacons to communicate among the MSs in the third strategy. If a sink $S_i$ is communicated by another sink $S_j$, it recomputes its movements (i.e. speed and direction) to avoid getting close to $S_j$, as below.
\begin{equation}
\textbf{v}^*_{S_i}=\textbf{t}+\textbf{v}_{S_i},
\end{equation}
where $\textbf{v}_{S_i}$ is the vector of the current velocity of $S_i$, and $\vert\textbf{t}\vert=\lambda\vert\textbf{v}_{S_i}\vert$.
\begin{align*}
\lambda=\left\{\begin{array}{l l}
\frac{R}{d}, &\quad \lceil\frac{d}{R}\rceil\leq \beta \\ \nonumber
0, &\quad  \text{otherwise} \nonumber
\end{array} \right.
\end{align*}
in which $d$ is the distance between $S_i$ and $S_j$, and $R$ is the communication range of the sinks. $\beta$ is the control parameter for the hop distance. In this strategy, though the sinks still move randomly but more evenly deploy in the sensing field.

\subsection{Multiple Sinks - Fixed/Predictable Mobility}
In the random mobility, movements of multiple sinks frequently causes topology changes in the network without prediction, which may lead to high packet loss rate. To address the issue, the MSs are designed to travel on fixed or predictable paths.

Let us start with a very simple but realistic model of the fixed mobility in a MWSN. In some applications such as observing traffic conditions or monitoring city environmental parameters, vehicles are employed as MSs, which normally traverse on predefined routes, to collect data from isolated sensor nodes deployed round a city. In the work \cite{konstantopoulos2012rendezvous}, a group of the sensor nodes which can communicate to each other is considered as a cluster, where a node closest to the sink path is elected as a cluster head. A new proposition in this work is that redundancy of spatial-temporal data collected by sensor nodes is filtered before the raw-processed information is transmitted to a sink, which can substantially reduce energy dissipation on data delivery. Likewise, Wang \textit{et al.} in \cite{wang2017energy} also exploited cluster basis but for a general network to prolong its longevity. The crucial idea in the work is that the MSs are expected to run along a border of the sensing field and pick up the sensor readings from the closest CH in a single hop communication. More importantly, the next location of a sink is dynamically computed on a principle whether the sink should be a next hop of the closest CH or not.\\
In other works, trajectories of MSs can be established from the predefined cluster structure. For instance, a two-phase communication protocol was proposed in \cite{gao2011efficient}, where topology information of a network is learned in the first phase, which is employed to arrange clusters and select cluster heads. An integer linear programming problem was then formulated to find the shortest paths for sink mobility, whose solutions can be obtained by the use of a genetic algorithm. Energy-efficient data collection is conducted in the second phase. In contrast to \cite{gao2011efficient}, Xing \textit{et al.} in \cite{xing2008rendezvous} assumed that sinks traverse on predefined trajectories of a routing tree. They then proposed two methods to optimize CHs or RPs. In the first technique, RPs are selected so that energy consumption in the network is minimize while a maximum ratio of network energy redundancy to sink travel distance total is considered to elect RPs in the second approach.
\begin{figure*}[tb]
	\centering
	\includegraphics[width=1\textwidth]{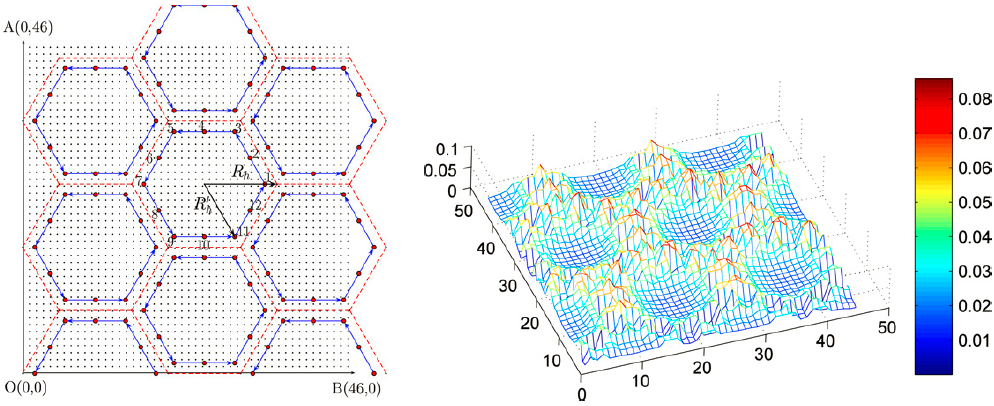}
	\caption{Mobile sinks move on edges of hexagons: Left - trajectories, Right - sensor energy dissipation in joules after 12 iterations \cite{marta2009improved} }\label{Figmultisink_fixed}
\end{figure*}
In a different context, the work \cite{marta2009improved} develop an algorithm that energy-effectively drives the sinks along perimeters of predefined hexagons. As demonstrated in Fig. \ref{Figmultisink_fixed}, the network energy is efficiently balanced among the sensor nodes. Furthermore, it proved that the proposed mechanism can extend the network working period up to 4.8 times as compared to that of a network with static sinks. In most of the examined works, the MSs are deemed to move in a free space. Nevertheless, in scenarios where there are obstacles in the field, how to drive the sinks to avoid collisions with obstacles while effectively collect all the sensor readings with shortest communications to the nodes? Xie \textit{et al.} \cite{xie2017energy} proposed a mechanism to divide the field into equal regions, where a sink is expected to stop at one point in each region to loading data from its nodes. If there happen obstacles in a region, none of MSs will travel into that region and all the sensor nodes in that region are required to transmit data to the sink in the closest obstacle-free region. The complete graph of the network can then be obtained by Warshall-Floyd technique, and the optimal paths for the MSs can be formulated as follows,

\begin{subequations}
\label{multisink_fixed_equ1}
\begin{alignat}{2}
&\textbf{min}        &\qquad& \sum_{i,j\in O, i\neq j} D(i,j)\gamma_{ij} \\
&\textit{subject to} &      & \sum_{i\in O, i\neq j}\gamma_{ij}=I_j, \;\; \forall j\in O\\
&	 &      & \sum_{j\in O, j\neq i}\gamma_{ij}=I_i, \;\; \forall i\in O\\
&	 &      & \sum_{j\in N(i,l)}I_i\geq 1, \;\; \forall i\in S\\
&	 &      & \sum_{j\in N(i,l)}x_{ij}^{(l)}-\sum_{k:i\in N(k,l)}x_{ki}^{(l)}=d_i, \;\; \forall i\in S,\forall l\in O\\
&	 &      & x_{ij}^{(l)}\geq 0, \;\; \forall i\in S, \forall l\in O, j\in N(i,l),
\end{alignat}
\end{subequations}
where $O$ and $S$ are sets of the sink locations on a path of avoiding obstacles and the sensor nodes, respectively. $N(i,l)=\{j\in S\cup\{l\}\mid D(i,j)\leq R, j\neq i\}$ is the neighbors of node $i$, and $D(i,j)$ is the distance between node $i$ and $j$ while $R$ is the communication range of a sensor.
\begin{align*}
\gamma=\left\{\begin{array}{l l}
1, &\quad \text{if} \; D(i,j)\; \text{on the sink path}  \\ \nonumber
0, &\quad  \text{otherwise} \nonumber
\end{array} \right.
\end{align*}

\begin{align*}
I_i=\left\{\begin{array}{l l}
1, &\quad \text{if location} \; l\in O\; \text{on the sink path}  \\ \nonumber
0, &\quad  \text{otherwise} \nonumber
\end{array} \right.
\end{align*}
$d_i$ denotes the node $i$ to be able to sense data from environment, and $x_{ij}$ is the rate assignment at from node $i$ to node $j$ when the sink at $l\in O$. The optimization problem (\ref{multisink_fixed_equ1}) is then solved by the spanning graph algorithm.

With an objective of decreasing power dissipation and increasing packet delivery ratio in a MWSN, the work \cite{inproceedingsWu} developed an interest dissemination with directional antenna strategy that enables the MSs to pre-inform CHs or RPs what interested data packets they want to collect, which sensor nodes would readily arrange for them when they arrive.

\subsection{Multiple Sinks - Controlled Mobility}
In addition to random or fixed mobility patterns, the MSs can be adaptively controlled to destinations with specified objectives such as coverage maintenance or energy efficiency \cite{natalizio2013controlled}. In this section, we will survey how the sinks are controllably driven.

\begin{figure*}[tb]
	\centering
	\includegraphics[width=0.8\textwidth]{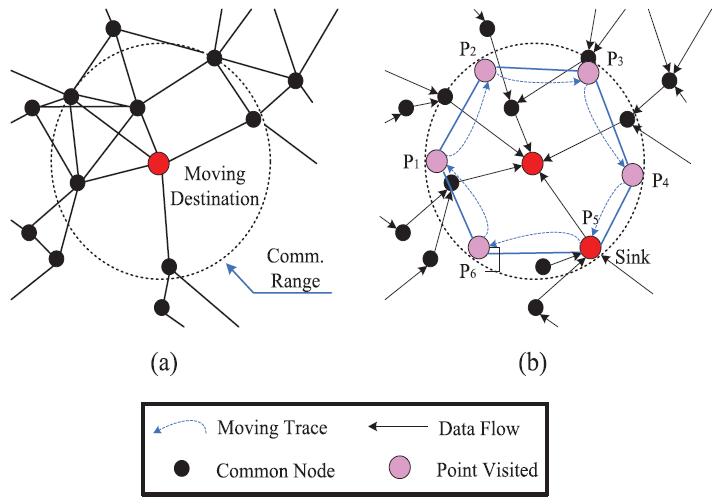}
	\caption{The sink selects a sojourn position where it can receive the most data packets from the Moving Destination \cite{Bi2007}}\label{multisink_control1}
\end{figure*}

In the first instance, mobility of the MSs in a MWSN is optimized along with deploying sensors, scheduling activity and routing data in a mixed integer linear programming optimization problem, which solutions are obtained by the practical heuristic algorithms \cite{Keskin2014}. To address this multi-variable optimization formulation, locations of sensors and schedules of activities are sequentially assumed to be constants in the first and second steps before movements of the MSs can be found. Another heuristic method was also proposed in \cite{Bi2007} to drive the MSs in each data gathering period. When the sinks know distribution of residual energy in the network, they automatically identify the highest residual energy nodes that are considered as moving destinations. Nonetheless, the sinks will not move to those sensor nodes but to locations of their communication ranges so that those highest residual energy nodes will be exploited to spend power in forwarding data for other nodes, which ultimately prolongs longevity of the network. Fig. \ref{multisink_control1} demonstrates one of the MSs can move to six locations around a moving destination for collecting data. Two other approaches based on simulated annealing \cite{Keskin2015} and Lagrangian \cite{Keskin2017} can also be employed to decide how the sinks can move in an energy-efficient data collection network. Though both the algorithms are heuristic, it was proved that the network lifetime can be considerably extended.

After the primitive control manner for the sink mobility in a MWSN with multiple modes of single, multiple and concast flows was discussed in \cite{goldenberg2004towards} to enhance energy-efficient performance in wireless communication, Wang \textit{et al.} in \cite{Wang2014} proposed novel routing protocols, which govern making decisions in the MS movements and lengthening the network longevity. More specifically, the authors employed the energy-aware routing maximum capacity path method \cite{Huang2004} to design the sink movement strategies. In the proposed algorithm, the sinks are expected to move to locations that can arrange communication ranges in the network based on residual energy levels of sensor nodes.
Some other authors exploited the cluster paradigm to design the network mobility. In \cite{banerjee2010increasing}, Banerjee \textit{et al.} proposed to employ the MSs as CHs, where each cluster is formed in a high residual energy sensor area and a CH is expected to move toward there for data transmission. The sink movements are obtained by heuristic algorithms in which collaborative talks among CHs are also effectively established. The work demonstrated that the proposed approach can elongate the network working time up to 75\%. In a similar fashion, to maintain effective connections among sinks and sensor nodes, a self-organizing and adaptive dynamic clustering algorithm was presented in \cite{Abuarqoub2017}, where the mobile collectors are driven to well-delimited clusters named service zones. In each service zone, the sink is moved to a collection zone so that signaling overhead, latency and bandwidth utilization are declined while network scalability and load balancing are increased. The evaluation performance showed 53\% increment in the network longevity. In a more strict way, the mobile collectors are required to visit the elected RPs within a allowed sojourn \cite{Xing2008}. In other words, the RPs can be selected among sensor nodes either when the mobile collector movements along the data routing tree or by ratios of residual network energy and MS travel distance.
In the context of path constraints, the sinks are controllably navigated through the network so that the far-off sensor nodes easily communicate with the others within the sink communication range, which eventually maximize data collected and minimize energy consumed by the network, as discussed in \cite{Gao2011}. To that end, the authors proposed a 0-1 integer linear programming maximum amount shortest path problem as follows,
\begin{subequations}
\label{multisink_control_equ1}
\begin{alignat}{2}
&\textbf{min}        &\qquad& \sum_{i=1}^n h_i \\
&\textit{subject to} &      & r_j\geq r_j^m \;\; \forall j=1,...,n_s \;\; \text{if} \;\; n_m\geq \sum_{i=1}^{n_s}r_i^m\\
&	 &      & \sum_{j=1}^{n_s}r_j=n_m,
\end{alignat}
\end{subequations}
where $h_i$ is the shortest hops from sensor $i$ to its destination sub-sink, a sensor node within direct communication from a sink. $n_s$ and $n_m$ are sub-sinks and sensor nodes faraway from a MS, respectively. While $r_j$ is the number of the far-off sensor nodes expected to communicate to the sub-sink $j$, $r_j^m$ is the limitation on $r_j$. The optimization problem is then resolved by a genetic two-dimensional binary chromosome algorithm. As an add-on to the path constraint problem in \cite{Gao2011}, Liang \textit{et al.} in \cite{Liang2011} constrained not only routing paths of sensor nodes but also traversing distances of sinks. Therefore, the constrained path problem becomes optimizing trajectories of the mobile collectors and their sojourn time at each location along their ways, which leads to substantially extended lifetime of the network. Though the proposed optimization is NP-hard, it can be effectively addressed by a three-stage heuristic algorithm, where it proved that the network working time can be prolonged up to 93\% as compared with the optimal one.

\begin{figure*}[tb]
	\centering
	\includegraphics[width=1\textwidth]{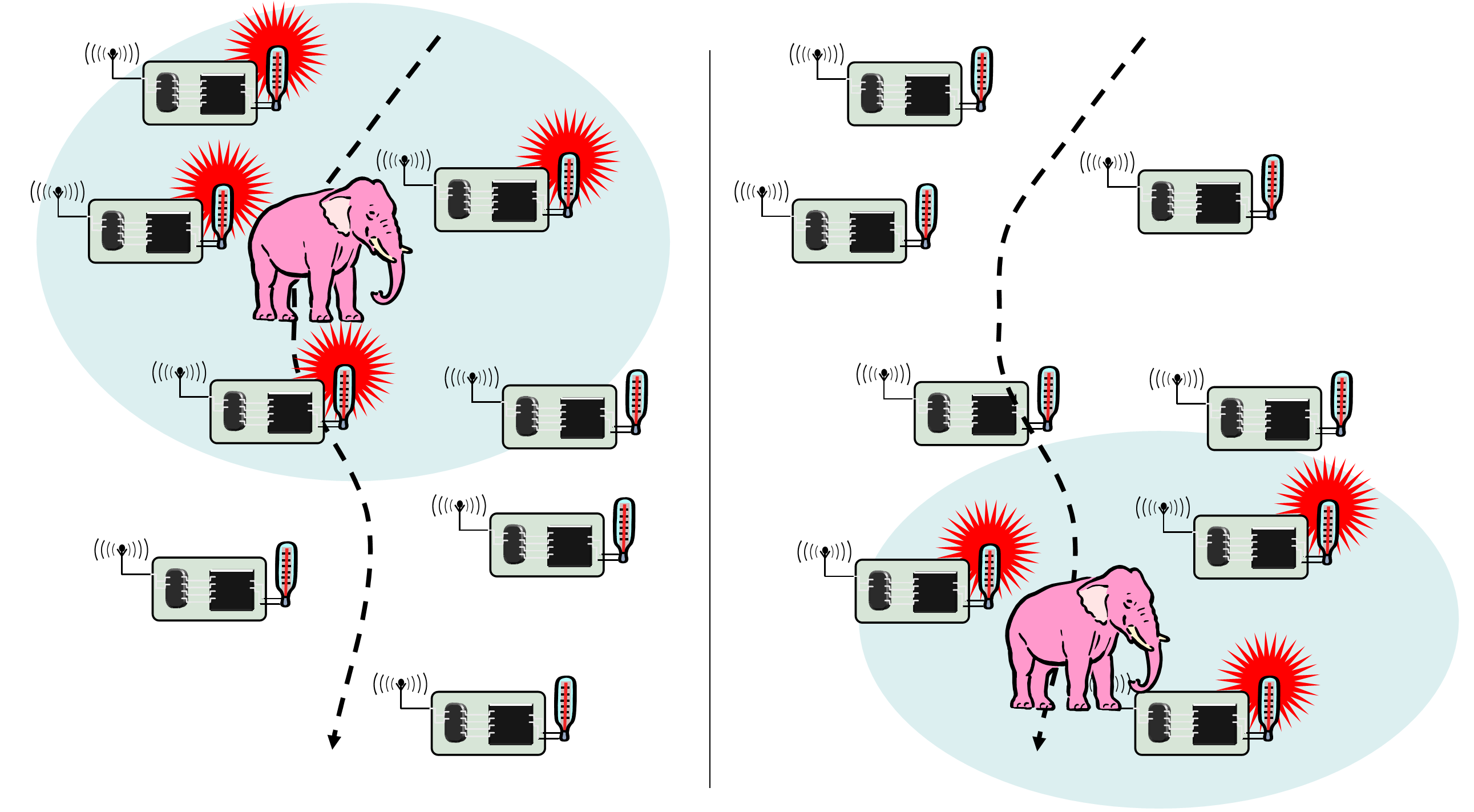}
	\caption{Event mobility type in \cite{karl2007protocols}  }\label{FigF3}
\end{figure*}

\begin{figure}[tb]
	\centering
	\includegraphics[width=0.45\textwidth]{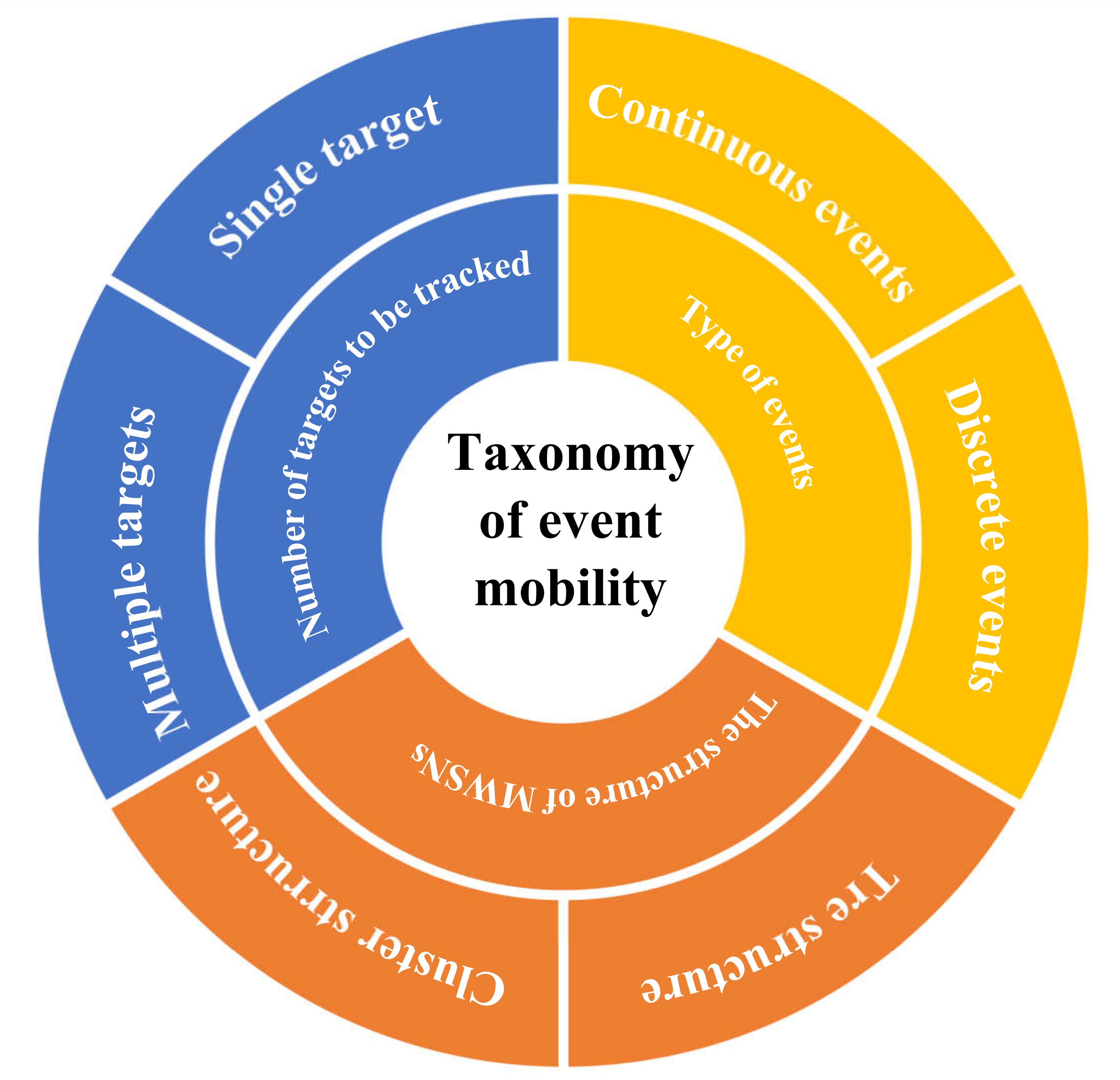}
	\caption{Taxonomy of event mobility }\label{Figeventmobility}
\end{figure}

\section{Event Mobility}
\label{sec_6}
In the previous sections we discussed the mobility patterns of both sensor nodes and sinks in a general MWSN, where data is continuously captured over time. Nevertheless, in the event-driven applications, where the data is merely generated as events happen, how is the network longevity improved? For instance, in a habitat monitoring application as shown in Fig. \ref{FigF3}, an animal is observed by a group of sensor nodes that are triggered to be awake by appearance of the animal. Nonetheless, they will then go to sleep mode once the animal disappears. 
The event mobility based approaches can be classified according to many aspects such as the number of events to be tracked, type of events or structure of MWSNs \cite{naderan2009mobile,ramya2012survey,ez2016comparative}. The classification is depicted in Fig. \ref{Figeventmobility}, where
\begin{itemize}
\item [$\bullet$] Number of events to be tracked:
	\begin{itemize}
	\item [$\star$] Single event: All sensor nodes in a network only track one event at the same time \cite{song2007cross,zhang2004dctc}.
	\item [$\star$] Multiple targets: More than one events can be tracked at the same time \cite{chen2004dynamic,lin2006efficient}.
	\end{itemize}
\item [$\bullet$] Type of events: Continuous events such as forest fires, oil spills or biochemical material \cite{chang2008coda,zha2004dynamic} and discrete events including people, animal or vehicles \cite{lin2006efficient,chen2004dynamic}.
\item [$\bullet$] Structure of MWSNs:
	\begin{itemize}
	\item [$\star$] Cluster structure: Sensor nodes are organized into clusters, where every cluster member senses events and then forwards data to its cluster head node \cite{chang2008coda,xu2004dual}. The structure of clusters can be static or dynamic where their structure can adaptively change due to absence of events.
	\item [$\star$] Tree structure: Sensor nodes are organized in a hierarchical tree \cite{lin2006efficient,zhang2004dctc}, where the event data is collected by leaf nodes then and forwarded to subtree nodes before reached to a root node.
	\end{itemize} 

\end{itemize}

In the work \cite{tan2010information}, the authors designed an information quality based mechanism for energy-efficient data transmission in an event-driven sensor network. The information quality in the network is defined by whether a target is detected. In the proposed method, a sensor node only captures information of a target when it appears nearby the sensor location before the information is transmitted to a base station through a pre-built distance-based aggregation tree. In other words, the sensor node is not active if there is no target around it, which leads to reduction of energy consumption and end-to-end delay over the network. In another work \cite{Chen2016}, it proposed to employ multiple MSs for the same application of the target detection. A new grid based data transmission paradigm was developed, where information of mobile targets is efficiently transmitted to the sinks through a grid tree that is constructed at the bottom grid cells. Data is also aggregated before sent, which considerably ameliorates not only the traffic flow but also the network lifetime.
 
\begin{figure*}[tb]
	\centering
	\includegraphics[width=0.8\textwidth]{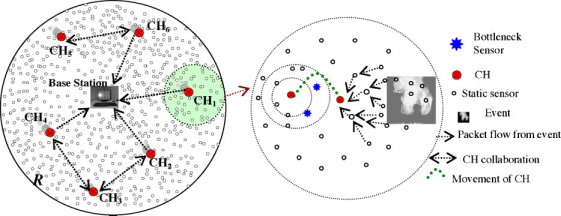}
	\caption{Event mobility type in \cite{banerjee2010increasing}}\label{event_1}
\end{figure*}

In other examples such as intrusion detection, when an intruder occurs, a number of mobile CHs may be sent toward the event region so that data transmission distances can be reduced, as shown in Fig. \ref{event_1}.

For instance, Ez-Zaidi \textit{et al.} in \cite{ez2015comparative} presented a new energy-efficient clustering approach to extend the network lifetime, where unexpected events including forest fire, intruder or randomly moving targets can be detected. In the proposed algorithm, the CH nodes are adaptively elected along trajectories of moving targets. In equivalent words, a sensor node is selected as a CH if it satisfies two conditions of high residual energy and close to the target. It is noticed that validity of a CH means there is appearance of a target. If the target disappears, all the sensor nodes play the same role. The dynamic CH election may reduce overhead and eliminate boundary problem, which results in energy efficiency in the network.

In a large-scale network, a distributed but energy-efficient data collection framework was proposed in \cite{sabbineni2010datacollection}. The method discretizes a sensing field into a grid and a MS is requested to gather data from some specific grid cells. By the use of a two-tier distributed hash table based technique each sensor node is aware of its location in a single cell. When an event occurs, sensor nodes within the event area will inform the network and the closest sink will move toward the event so that data can be gathered in a single hop, which can not only save sensor energy but also reduce execution time of data transmission operation. In order to correctly drive a sink or a mobile CH to the event region, \cite{banerjee2010increasing} formulated the event center, where its 2D coordinates ($x_c$, $y_c$) can be computed as follows,
\begin{equation}
x_c=\frac{\sum_{i\in s_{j}}(\vert x_{\text{CH}_j}-x_i)\times g(i)}{\sum_{i\in s_{j}}g(i)},
\end{equation}
\begin{equation}
y_c=\frac{\sum_{i\in s_{j}}(\vert y_{\text{CH}_j}-y_i)\times g(i)}{\sum_{i\in s_{j}}g(i)},
\end{equation}
where $s_{j}$ is the set of sensors in a cluster with $\text{CH}_j$ while ($x_i$, $y_i$) and ($x_{\text{CH}_j}$, $y_{\text{CH}_j}$) are coordinates of the $i^{th}$ sensor and $\text{CH}_j$. $g(i)$ is the transmission speed of the $i^{th}$ sensor. A MS is navigated to this center point for single-hop data dissemination.

In the context of optimization, a trajectory of a mobile CH or sink in an event monitoring network can be obtained under constraint of data collection deadline. For instance, Tashtarian \textit{et al.} in \cite{tashtarian2015odt} formulated the sink trajectory optimization problem by incorporating factors including the number of sensors actively monitoring the event, the sink velocity, sojourn time and deadline to capture information. Though the proposed optimization problem is NP-hard, it can energy-efficiently addressed by the use of a decision tree and the dynamic programming approach. In other words, the sensing field is divided into small autonomous zones, whose sizes are small enough for the sink to communicate with sensor nodes in a single-hop communication range. Furthermore, in a scenario, where a future location of an event can be forecasted by utilizing its historical data, e.g. intruder's movements, based on the network information, the sink can effectively calculate its next optimal location \cite{Vincze2007}. In fact, the next movement of the sink must minimize either total energy consumed in the whole network or maximum energy load on a single sensor in that step. The results obtained in simulations have shown that the network can be prolonged up to 150\% as compared to one without mobile capability.

\section{Conclusions and Future Directions}
\label{sec_7}
In this paper, we have presented a comprehensive survey on methods proposed to exploit mobility of sensor nodes and sinks to prolong the network longevity of a MWSN. After a brief introduction of the frequently utilized definitions of the network lifetime for convenient reading, we have summarized the strategies employing the sensor node mobility to efficiently utilize power over the network so that it can last in the longest period of time possible. We have then comprehensively reviewed the techniques proposed for the MWSN given its sink mobility with the purpose of maximizing the fully functional operations in the network. Those approaches have been systematically classified, depending on the mobility patterns they design for the sink(s) including random, fixed/predictable and controlled modalities. In the cases where the events move and arbitrarily appear, we have also surveyed the researched algorithms proposed for a MSWN so that its energy utilization is maximized.

A MWSN can be considered as an important part of an IoT system; hence future works are expected to effectively exploit mobile platforms that are already available in the IoT infrastructure to significantly ameliorate the longevity of the battery constrained MWSNs.


%


\ifCLASSOPTIONcaptionsoff
  \newpage
\fi

\balance

\bibliographystyle{IEEEtran}
\bibliography{References}


\end{document}